\documentclass[twocolumn]{aastex63}

\usepackage{amsmath}
\usepackage{multirow}
\usepackage{hyperref}

\accepted{}
\shorttitle{Lifetimes of protoplanetary disks}
\shortauthors{Ronco et al.}
\graphicspath{{./}{figures/}}

\begin{document}



 

\title{Long live the disk: lifetimes of protoplanetary disks in hierarchical triple star systems \\
and a possible explanation for HD 98800~B}

\correspondingauthor{Mar\'{\i}a Paula Ronco}
\email{mronco@astro.puc.cl}

\author[0000-0003-1385-0373]{Mar\'{\i}a Paula Ronco}
\affil{Instituto de Astrof\'{\i}sica - Pontificia Universidad Cat\'olica de Chile,
Av. Vicu\~na Mackenna 4860, Macul - Santiago, 8970117 , Chile.}
\affil{N\'ucleo Milenio de Formaci\'on Planetaria (NPF), Chile.}

\author[0000-0001-8577-9532]{Octavio M. Guilera}
\affil{Instituto de Astrof\'{\i}sica de La Plata, CCT La Plata-CONICET-UNLP, Paseo del Bosque S/N (1900), La Plata, Argentina.}
\affil{Instituto de Astrof\'{\i}sica - Pontificia Universidad Cat\'olica de Chile,
Av. Vicu\~na Mackenna 4860, Macul - Santiago, 8970117 , Chile.}
\affil{N\'ucleo Milenio de Formaci\'on Planetaria (NPF), Chile.}

\author[0000-0003-1965-3346]{Jorge Cuadra}
\affil{Departamento de Ciencias, Facultad de Artes Liberales, Universidad Adolfo Ib\'a\~nez, Avenida Padre Hurtado 750, Vi\~na del Mar, Chile.}
\affil{N\'ucleo Milenio de Formaci\'on Planetaria (NPF), Chile.}

\author[0000-0001-8031-1957]{Marcelo M. Miller Bertolami}
\affil{Instituto de Astrof\'{\i}sica de La Plata, CCT La Plata-CONICET-UNLP, Paseo del Bosque S/N (1900), La Plata, Argentina.}
\affil{Facultad de Ciencias Astron\'omicas y Geof\'{\i}sicas, Universidad Nacional de La Plata, Paseo del Bosque S/N (1900), La Plata, Argentina.}

\author[0000-0003-3713-8073]{Nicol\'as Cuello}
\affil{Univ. Grenoble Alpes, CNRS, IPAG (UMR 5274), F-38000 Grenoble, France}

\author{Camilo Fontecilla}
\affil{Instituto de Astrof\'{\i}sica - Pontificia Universidad Cat\'olica de Chile,
Av. Vicu\~na Mackenna 4860, Macul - Santiago, 8970117 , Chile.}
\affil{N\'ucleo Milenio de Formaci\'on Planetaria (NPF), Chile.}

\author{Pedro Poblete}
\affil{Astrophysikalisches Institut, Friedrich-Schiller-Universität Jena, Schillergäßchen 2–3, 07745 Jena, Germany}

\author[0000-0001-7868-7031]{Amelia Bayo}
\affil{Instituto de F\'isica y Astronom\'ia, Facultad de Ciencias, Universidad de Valpara\'iso, Chile}
\affil{N\'ucleo Milenio de Formaci\'on Planetaria (NPF), Chile.}

\nocollaboration{8}



\begin{abstract}
The gas dissipation from a protoplanetary disk is one of the key processes affecting planet formation, and it is widely accepted that it happens on timescales of a few million years for disks around single stars. Over the last years, several protoplanetary disks have been discovered in multiple star systems, and despite the complex environment in which they find themselves, some of them seem to be quite old, a situation that may favor planet formation. 
A clear example of this is the disk around HD 98800~B, a binary in a hierarchical quadruple stellar system, which at a $\sim$10~Myr age seems to still be holding significant amounts of gas. Here we present a 1D+1D model to compute the vertical structure and gas evolution of circumbinary disks in hierarchical triple star systems considering different stellar and disk parameters.

We show that tidal torques due to the inner binary together with the truncation of the disk due to the external companion strongly reduce the viscous accretion and expansion of the disk. Even allowing viscous accretion by tidal streams, disks in these kind of environments can survive for more than 10 Myr, depending on their properties, with photoevaporation being the main gas dissipation mechanism. We particularly apply our model to the circumbinary disk around HD 98800~B and confirm that its longevity, along with the current non-existence of a disk around the companion binary HD 98800~A, can be explained with our model and by this mechanism.  
\end{abstract}



\section{Introduction} 
\label{sec:intro}

Stellar multiplicity is the most common outcome of star formation \citep{DucheneKrauss2013,Bate2018} and it has been shown that 
the younger a stellar object is, the greater the probability that is part of a multiple stellar system \citep{Reipurth2014, Tobin2016}. 

As a by-product of their formation, young stellar objects are surrounded by disks of gas and dust called protoplanetary disks, in where planet formation begins. Since the spectacular image of HL Tau \citep{ALMA2015} more than five years ago, the number of resolved disks around stellar objects of different kinds and ages has grown significantly. But protoplanetary disks have been found not only around single stars. Several surveys discovered many of them in binaries and even in higher multiplicity star systems \citep{Correia2006, Cox2017, Cieza2019, Manara2019, Akeson2019, Zurlo2020}. 

Some of the most recent and spectacular examples are the ``cosmic pretzel" disk around the young [BHB2007] 11 binary system \citep{Alves2019}, the old polar circumbinary disk in the hierarchical quadruple star system HD 98800 \citep{Kennedy2019}, and the spectacular disks imaged in the triple-star system GW Orionis \citep{Kraus2020}. 

The evolution and lifetimes of disks in binary or multiple star systems are quite different from those of a protoplanetary disk around a single star.
Circumbinary disks in close-in binaries are especially affected in the inner regions, which are cleared out as a competition between the torques exerted by the inner binary that push the disk farther out, and the viscous evolution trying to close this cavity \citep{ArtymowiczLubow1994,ArtymowiczLubow1996, DelValle2012}. The size of the cavity is estimated to be between 2 and 5 times the inner binary separation,  increasing with the binary eccentricity \citep{ArtymowiczLubow1994,Thun2017,Hirsh2020, Ragusa2020}.

On the other hand, circumstellar disks orbiting only one of the stars in a binary system, are truncated by the tidal effect produced by the outer star \citep{ArtymowiczLubow1994}. This truncation imposes a zero-flux edge condition at a distance from the central star estimated to be a function of the mass ratio and separation of the binary \citep{PapaloizouPringle1977, Pichardo2005}. 

For circumstellar disks, a mechanism that significantly contributes to a faster disk dispersal is photoevaporation. This phenomenon is produced by the central star \citep{Alexander.et.al.2006, Gorti2009} but also from possible external sources if the environment presents OB stars \citep{Clarke2007, Anderson2013, winter2020}, or the star suffers close encounters \citep{Winter2018}.
The resulting circumstellar disk lifetimes are tipically below 10 Myr \citep{Fedele2010, Pfalzner2014}. These effects also influence disk evolution in multiple stellar systems.
In particular, circumbinary disks are expected to live longer than circumstellar disks \citep{Alexander2012}, but truncated circumstellar disks in wide binaries could dissipate faster \citep{Cieza2009, Kraus2012}. In this context, two questions are still under debate: i) how does photoevaporation affect disc evolution in hierarchical triple stellar systems, and ii) how do the above-mentioned processes affect planet formation altogether? 

Despite the complex dynamical environment where disks in binaries and triples form and evolve, planets can still form in them \citep{MarzariThebault2019}. Indeed, more than 150 exoplanets have been discovered around binary star systems and more than 30 in multiple star systems of different kinds \citep{Schwarz2016}.
Moreover, it has also been suggested that the frequency of planets in binary systems, particularly gas giants, can be as high as around single stars, except in the case of very close-in binaries \citep{Martin2018, BonavitaDesidera2020}. 

Hydrodynamical simulations play a key role
in modeling with great detail the gas and dust evolution in protoplanetary disks. However, the high computational cost that these type of simulations require makes not yet feasible to follow the whole disk evolution. For the moment, and despite their simplifications, 1D models close this gap, allowing us to study the evolution of protoplanetary disks until they completely dissipate. 

In this work we focus on describing the time evolution of protoplanetary disks in hierarchical triple star systems with a new 1D+1D model that considers viscous accretion, irradiation and X-ray photoevaporation from the inner binary. Our general aim is to compare their evolution with their circumbinary counterparts, and determine their global characteristics and dissipation timescales, for different stellar parameters.

A specific goal of this work is to understand the longevity of disks in hierarchical stellar systems. We then apply our model to the particular case of the disk in the hierarchical quadruple star system HD 98800, which constitutes the ideal astrophysical testbed to do it, as it is uncertain how a nearly 10 Myr old disk still has significant amounts of gas \citep{Ribas2018}. 

This paper is organized as follows. In sec.~\ref{sec:model} we describe our model for the evolution of disks in triple star systems. We then define our general setup of initial conditions in sec.~\ref{sec:ini_cond}. We describe our general results in sec \ref{sec:general_results} including a comparison between triple-star and binary-star disk evolution in sec.~\ref{sec:T1-vs-B1}, and a description of the stellar parameters dependency in sec.~\ref{sec:stellar_dependency}. In sec.~\ref{sec:HD98800B} we describe the particular case of the system HD 98800, and apply our model to the disk around HD 98800~B.  We present a discussion in sec.~\ref{sec:discussion} and conclude with some key aspects of our findings in sec.~\ref{sec:conclusions}.

\section{Model for the gas disk evolution in a hierarchical triple star system} \label{sec:model}

Here we present {\scriptsize{PLANETALP-B}}, a 1D+1D code that computes the time evolution of a gaseous disk in a hierarchical triple star system. {\scriptsize{PLANETALP-B}} is an extension of our code {\scriptsize{PLANETALP}} \citep{Ronco+2017, Guilera2017, Guilera2019, Guilera2020, venturini2020SE, Venturini2020}, a global model of disk evolution and planet formation that computes the growth of planets immersed in a circumstellar protoplanetary disk that evolves in time. This first version of {\scriptsize{PLANETALP-B}}, which can also be applied for the study of circumbinary disks and circumstellar disks around one of the stars in a wide binary star system, only focuses on their gaseous component evolution.

\subsection{The vertical disk structure}
 \label{subsec:vertical_structure}

\begin{figure}
  \centering
    \includegraphics[angle= 0, width= 0.95\columnwidth]{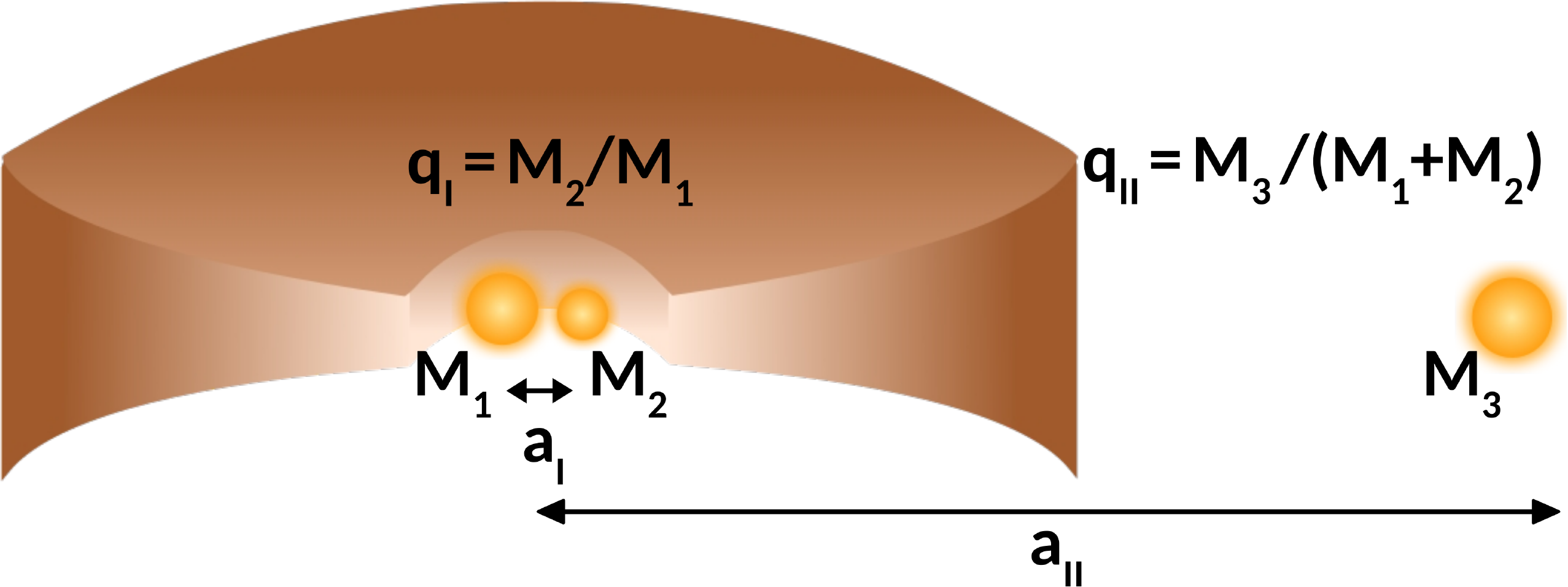}
    \caption{Schematic view of the configuration of our scenario of study: an axisymmetric circumbinary disk in a circular and coplanar hierarchical triple-star system.}
  \label{fig:Scheme}
\end{figure}

Our subject of study is a disk in a hierarchical triple star system, this is, a circumbinary disk affected by an external stellar companion. Figure \ref{fig:Scheme} shows an schematic view of this configuration which, due to the axi-symmetrical nature of our model, will always consider circular and coplanar orbits for the inner binary, the circumbinary disk and the outer stellar companion. Moreover, despite the disk being subject to a time dependent gravitational potential due to the stellar orbital motions, we consider the disk to be rotating in a gravitational potential given by the total mass of the inner binary system, and neglecting the gravitational perturbations of the external companion. 
We will discuss the limitations of these considerations in sec.~\ref{sec:discussion}. 

To compute the vertical structure of a disk in this environment we assume an axi-symmetric, irradiated disk in hydrostatic equilibrium. We follow \citet{Guilera2017, Guilera2019, Guilera2020} which use the same methodology described in \citet{Alibert2005,Migaszewski2015} for a circumstellar case, but considering extra tidal heating terms in the energy equation due to the effects of the inner binary and the external companion \citep{Lodato2009, Fontecilla2019}. The complete set of equations is given by:
 \begin{eqnarray}
  \frac{\partial P}{\partial z} &=& -\rho \Omega^2 z, \label{eq:Presion}\\
    {\frac{\partial F}{\partial z}} &=& \frac{9}{4} \rho \nu \Omega^2 + D_{\Lambda_{\text{I}}} + D_{\Lambda_{\text{II}}}, \label{eq:Flujo} \\
    \frac{\partial T}{\partial z} &=& \nabla \frac{T}{P}\frac{\partial P}{\partial z},
 \label{eq:Temperatura} 
\end{eqnarray}
where $P$, $\rho$, $F$, $T$ and $z$ represent the pressure, density, radiative heat flux, temperature and vertical coordinate of the disk, respectively. $\nu= \alpha c_s^2/\Omega$ is the viscosity \citep{Shakura1973}, where $\alpha$ is a dimensionless parameter, $c_s^2= P/\rho$ is the square of the locally isothermal sound speed,  $\Omega = \sqrt{GM_{\text{I}}/R^3}$ is the Keplerian frequency at a given radial distance $R$ from the central binary, and $M_{\text{I}}=M_1+M_2$ is its mass. We consider that the heat in the circumbinary disk is vertically transported by radiation or convection following the standard Schwarzschild criterion, as described in \citet{Guilera2019}.

The terms $D_{\Lambda_{\text{I}}}=(\Omega_{\text{I}} - \Omega)\Lambda_{\text{I}}\rho$ and $D_{\Lambda_{\text{II}}}=(\Omega_{\text{II}} - \Omega)\Lambda_{\text{II}}\rho$ in eq. \ref{eq:Flujo} represent the tidal heating dissipated in the form of radiation due to the inner binary and the outer stellar companion, respectively. 
$\Omega_\text{I}$ is the Keplerian frequency at the inner binary separation, $a_{\text{I}}$, given by $\Omega_{\text{I}} = \sqrt{G(M_{\text{I}})/a_{\text{I}}^3}$, and $\Omega_{\text{II}}$ is the Keplerian frequency at the separation between the inner binary system and the external companion star, $a_{\text{II}}$, computed as $\Omega_{\text{II}} = \sqrt{G(M_{\text{II}})/a_{\text{II}}^3}$, where $M_{\text{II}} = (M_1+M_2)+M_3$, the total mass of the stellar system. 

The quantities $\Lambda_{\text{I}}$ and $\Lambda_{\text{II}}$ represent the torques due to the inner binary and outer stellar companion, respectively. 
The classical expression for $\Lambda$ proposed by \citet{ArmitageNatarajan2002}\footnote{For a detailed discussion on this prescription we refer the reader to \citet{PetrovichRafikov2012, RafikovPetrovich2012, Rafikov2016}.
} 
for $q\ll1$ could be overestimating the tidal torque contribution on the disks outside the Lindblad resonances, which may modify the surface density profiles \citep{TazzariLodato2015}. Thus, following \citet{TazzariLodato2015, Fontecilla2019} we consider an exponential cutoff for the tidal torques outside this region, computed as:
\begin{eqnarray}
\Lambda_{\text{I}}(R) &=& \frac{f}{2}q_{\text{I}}^2\Omega^2R^2\left(\frac{a_{\text{I}}}{\Delta_{\text{I}}}\right)^4 e^{-\left(\frac{R-R_{\text{OM}}}{W_{\text{OM}}}\right)^2}, \label{eq:TorqueI} \\
\nonumber \\
\Lambda_{\text{II}}(R) &=& -\frac{f}{2}q_{\text{II}}^2\Omega^2R^2\left(\frac{R}{\Delta_{\text{II}}}\right)^4 e^{-\left(\frac{R-R_{\text{IM}}}{W_{\text{IM}}}\right)^2},
\label{eq:TorqueII}
\end{eqnarray}
where $f$ is a dimensionless normalization parameter that can adopt different values between 0.001 and 1, \citep{ArmitageNatarajan2002, Alexander2012, Vartanyan2016, 2Shadmehri2018}. 
The variable $q_{\text{I}}$ represents the mass ratio between $M_2$ and $M_1$, and $q_{\text{II}}$ the mass ratio between $M_3$ and $(M_1+M_2)$. Finally, $\Delta_{\text{I}} = {\text{max}}(R^{\text{I}}_{\text{Hill}},{\text{H}}_{\text{g}},|R-a_{\text{I}}|)$ and $\Delta_{\text{II}} = {\text{max}}(R^{\text{II}}_{\text{Hill}},{\text{H}}_{\text{g}},|R-a_{\text{II}}|)$, 
with $R^{\text{I}}_{\text{Hill}}=a_{\text{I}}(q_{\text{I}}/3)^{1/3}$ the Hill radius of the secondary star in the binary system,  $R^{\text{II}}_{\text{Hill}}=a_{\text{II}}(q_{\text{II}}/3)^{1/3}$ is the Hill radius of the third star and ${\text{H}}_{\text{g}}$ is the height scale of the disk. As in \citet{Fontecilla2019} we consider $R_{\text{OM}}=1.59a_{\text{I}}$ and $R_{\text{IM}}=0.63a_{\text{II}}$ being the radii of the outermost and innermost Lindblad resonances,  and $W_{\text{OM}}=75{\text{H}}_{\text{g}}$ and $W_{\text{IM}}=370{\text{H}}_{\text{g}}$ being the widths of the Gaussian smoothing. \\

Note that, if the case of study is that of a circumstellar disk,  $\Lambda_{\text{I}} = \Lambda_{\text{II}} = 0$, while if it is a circumbinary disk, only $\Lambda_{\text{II}} = 0$, and if it is a circumstellar disk affected by an external star companion, only $\Lambda_{\text{I}} = 0$.

The boundary conditions at the surface of the disk $H$, $P_s= P(z=H)$, $F_s= F(z=H)$, $T_s= T(z=H)$, considering the effects of the inner binary and outer star, are given by
\begin{eqnarray}
  P_s &=& \frac{\Omega^2 H \tau_{\text{ab}}}{\kappa_s}, \label{eq:CB1}\\
  F_s &=& \frac{3 \dot{M}_{\text{st}} \Omega^2}{8\pi} + \nonumber \\
  & & \frac{((\Omega_{\text{I}} -\Omega)\Lambda_{\text{I}} +  (\Omega_{\text{II}} -\Omega)\Lambda_{\text{II}})\dot{M}_{\text{st}}\Omega\mu m_{\text{H}}}{6\pi\alpha k_{\text{B}}(T^4_{\text{irr}}+T^4_s)^{1/4}},\label{eq:CB2}\\
  0 &=& 2 \sigma \left( T_s^4 + T_{\text{irr}}^4 - T_{\text{b}}^4 \right) - \frac{9 \alpha \Omega k (T_s^4 + T_{\text{irr}}^4)^{1/4}}{8 \kappa_s \mu m_{\text{H}}} - \nonumber \\
 & &  F_s - (\Omega_{\text{I}}-\Omega)\dfrac{\Lambda_{\text{I}}}{2\kappa_s} - (\Omega_{\text{II}}-\Omega)\dfrac{\Lambda_{\text{II}}}{2\kappa_s},
\label{eq:CB4}
 \end{eqnarray}
 
where $\tau_{\text{ab}}= 0.01$ is the optical depth, $T_{\text{b}}=
10$~K is the background temperature, and $\dot{M}_{\text{st}}$ is the
equilibrium accretion rate. In addition to the surface boundary conditions, due to symmetry we consider that the heat flux in the midplane of the disk should be $F(z=0)=F_0=0$. 

Since we now have an inner binary system, the temperature associated with the stellar irradiation is considered as 
\begin{equation}
  T^4_{\text{irr}}= T^4_{\text{irr}_1} + T^4_{\text{irr}_2},
  \label{eq:Irrad1} 
\end{equation}
i.e. we are adding up the fluxes due to each star, making the approximation that both are located at their centre of mass, where
\begin{eqnarray}
  T_{\text{irr}_i} &=& T_i \bigg[ \frac{2}{3\pi} \left( \frac{R_{\star_i}}{R} \right)^3 + \frac{1}{2} \left(\frac{R_{\star_i}}{R} \right)^2 \times \nonumber \\
  & & \left( \frac{H}{R} \right)
  \left( \frac{d\log H}{d\log R} - 1 \right) \bigg]^{0.5}, \text{i} = 1,2,   \label{eq:Irrad2} 
\end{eqnarray}
and $R_{\star_i}$ and $T_i$ are the radius
and effective temperature of the primary ($i=1$) and the secondary ($i=2$) star of the inner binary system. $R$ is the radial
coordinate, and $d\log H/d\log R= 9/7$. The individual stellar masses, with their corresponding stellar radii and temperatures, are obtained from \citet{Baraffe2015}. Note here that our model does not consider the possible effect of self-shadowing if the disk becomes non-flared nor the possible irradiation from the third star. 

Eq. \ref{eq:Presion} to \ref{eq:Temperatura} of the vertical structure are computed using a shooting method with a Runge-Kutta-Fehlberg integrator, while eq. \ref{eq:CB1} to eq. \ref{eq:CB4} are solved by a multidimensional Newton-Raphson algorithm. Solving these equations allows us to obtain the mean viscosity  $\overline{\nu}(\Sigma_{\text{g}},R)$, where $\Sigma_{\text{g}}$ is the gas surface density, required to solve the radial evolution of the disk.

It is worth noting that the detailed computation of the vertical structure and the thermodynamical quantities of circumbinary disks in hierarchical triple star systems can be particularly useful for future transfer radiative calculations.

\subsection{Time evolution of the radial disk structure}
\label{subsec:radial_structure}
In order to compute the time evolution of the gas surface density, we solve the classical 1D diffusion equation \citep{Pringle1981} including the terms corresponding to the torques injected by the binaries, 
already described in the previous subsection,
\begin{align}
  \frac{\partial \Sigma_{\text{g}}}{\partial t}= & \frac{3}{R}\frac{\partial}{\partial R} \left[ R^{1/2} \frac{\partial}{\partial R} \left( \overline{\nu} \Sigma_{\text{g}} R^{1/2}  \right) - \frac{2\Sigma (\Lambda_{\text{I}} + \Lambda_{\text{II}})}{3\Omega} \right] + \nonumber \\  
  & \dot{\Sigma}_{\text{w}}(R), 
\label{eq:evol_gas}
\end{align}
where $t$ is the temporal coordinate, $\overline{\nu}$ the mean viscosity at the mid-plane of the disk, $\Sigma_{\text{g}}$ the gas surface density and $\dot{\Sigma}_{\text{w}}(R)$ the sink term due to the X-ray photoevaporation.  The latter is computed  following the analytical prescriptions derived by \citet{Owen2012} from radiation-hydrodynamic models \citep{Owen2011,Owen2012}. The total mass loss rate depends on the X-ray luminosity of the central star, which in our case is actually a binary.  Therefore, we use 
\begin{equation}
L_{\text{X}} = L^1_{\text{X}} + L^2_{\text{X}}
\label{eq:XrayL}
\end{equation}
where $L^1_{\text{X}}$ and $L^2_{\text{X}}$ are the X-ray luminosities of the stars of the inner binary system, which can be computed following \citet{Preibisch2005},
\begin{eqnarray}
    \text{log}&(&L^i_{\text{X}}[{\text{erg~}}{\text{s}}^{-1}]) = 30.37 + \nonumber \\
    & & 1.44\log\left(\frac{M_i}{M_\odot}\right), \text{i}=1,2,
    \label{eq:Lx}
\end{eqnarray}
who derived this correlation between X-ray luminosity and stellar mass for young ($5<\log\tau [\text{yr}] <9.5$) low-mass stars $(M_\star \leq 2M_\odot)$. 
Then, as described in Appendix B of \citet{Owen2012}, for primordial disks (or disks without holes) the total mass-loss rate is given by
\begin{eqnarray}
    \dot{M}_{\text{w}} &=& 6.25\times10^{-9}\left(\frac{M_{\text{I}}}{1M_\odot}\right)^{-0.068} \times \nonumber \\
    & & \left(\frac{L_\text{X}}{10^{30}\text{erg~}\text{s}^{-1}}\right)^{1.14}~M_\odot\text{yr}^{-1},
    \label{eq:Mdot_disk}
\end{eqnarray}
while for disks with holes or transition disks is given by
\begin{eqnarray}
    \dot{M}_{\text{w}} &=& 4.8\times10^{-9}\left(\frac{M_{\text{I}}}{1M_\odot}\right)^{-0.148} \times \nonumber \\
    & & \left(\frac{L_{\text{X}}}{10^{30}{\text{erg~}}{\text{s}}^{-1}}\right)^{1.14}~M_\odot{\text{yr}}^{-1}.
    \label{eq:Mdot_hole}
\end{eqnarray}
These approximations were derived for circumstellar disks which can present both kinds of morphologies at different epochs. However, circumbinary disks always present inner cavities and thus, we only compute the X-ray mass loss rate following eq. \ref{eq:Mdot_hole}. 

From the normalised radial mass-loss profiles derived by \citet{Owen2012} and presented in their Appendix B, and by computing $\dot{M}_{\text{w}}$ (from eq. \ref{eq:Mdot_hole} in our case),  we can compute $\dot{\Sigma}_{\text{w}}(R)$ that verifies
\begin{equation}
\dot{M}_{\text{w}} = \int 2\pi r \dot{\Sigma}_{\text{w}}(R)dR.
\end{equation}

As in \citet{RosottiClarke2018}, we are not considering the possible contribution from the external companion to the photoevaporation rate of the circumbinary disk. The choice to only consider X-ray photoevaporation in our model is due to the fact that this type of irradiation has yielded much higher photoevaporation rates than the EUV rates \citep{ErcolanoOwen2010, Owen2011, Picogna2019}. Moreover, \citet{Kunitomo2021} recently showed that this is the predominant disk dispersal mechanism for low-mass stars ($<2.5M_\odot$), compared to EUV and FUV photoevaporation (see their Fig. 11).

An important point to mention here is that, in the 1D approach, the form of the eqs. \ref{eq:TorqueI} and \ref{eq:TorqueII} completely prevents viscous gas accretion onto the stars. However, as shown by \citet{ArtymowiczLubow1996} and many others \citep[e.g.,][]{Cuadra2009, Dunhill2015, Tang2017},  accretion onto the stars still happens via tidal streams, a mechanism that can only be captured in 2D and 3D simulations. This mass loss rate is not constant but modulated by the binary orbit. While \citet{MacFadyen2008}
estimated that this accretion could be in average $\sim 10\%$ of the one expected in the absence of the binary, \citet{DOrazio2013, Farris2014} suggested that it could reach up to 30-60$\%$. The results of \citet{Ragusa2016} are in agreement with these previous works when the disk is thick. However, these authors find that it is possible to suppress all mass accretion if the disk becomes sufficiently thin. Due to the uncertainties on the amount of this mass loss, our model allows that a certain fraction $\epsilon$ of the expected 
accretion rate is accreted by the inner binary. In a similar way as \citet{Alexander2012} did, we compute the 
accretion rate $\dot{M}_{\text{B}}$ as $\dot{M}_{\text{B}}=3\pi\nu\Sigma$ at every radial bin between $R_{\text{c}}$ and $2R_{\text{c}}$ 
(where $R_{\text{c}}$ is the radius of the inner cavity), and adopt the maximum value of $\dot{M}_{\text{B}}$ in that region, considering it as an upper limit. 

The disk radial structure is solved using an implicit Crank-Nicholson method, from $a_{\text{I}}$ to $a_{\text{II}}$ using 2000 radial bins logarithmically
equally spaced. We set boundary conditions of zero torque, which is equivalent to having zero density at the boundaries. We note here that the boundary conditions are applied to the radial bins corresponding to $a_{\text{I}}$ and $a_{\text{II}}$. However, as we show in the next sections, the tidal torques included in the model naturally truncate the disk at different locations. 
A validation of our 1D+1D model for the case of circumbinary disks is presented in appendix A, comparing our results to those of \citet{Vartanyan2016}.

\section{Initial Conditions and general Setup}
\label{sec:ini_cond}

The initial gas surface density profile for our protoplanetary disk is given by 
\begin{eqnarray}
  \Sigma_{\text{g}} &=& \Sigma_{\text{g}}^0 \left( \frac{R}{R_0} \right)^{-\gamma} e^{-(R/R_0)^{2-\gamma}}, \label{eq1-sec2-0}
\end{eqnarray}
where $R_0$ is the characteristic radius of the disk and $\gamma$ the gas surface density exponent.  $\Sigma_{\text{g}}^0$ is a normalization parameter that depends on the disk mass.

The inner and outer radii of our circumbinary disk are set initially equal to the star separations, $a_{\text{I}}$ (inner binary) and $a_{\text{II}}$ (outer binary). However, the inner and outer radii of the disk will naturally evolve (expand and contract, respectively) as a consequence of the tidal torques induced by the inner binary and the external star (represented by the term that includes $\Lambda_{\text{I}}$ and $\Lambda_{\text{II}}$ in eq. \ref{eq:evol_gas}). 

We define a general setup of initial conditions, in where we fix the protoplanetary disk parameters and only vary the stellar ones.
As shown in Table \ref{tab:CI-Generales}, we perform simulations for a fixed disk of $M_{\text{d}}=0.05M_\odot$, with $R_0=39$~au, $\gamma=1$,\footnote{These are typical values obtained for circumstellar disks \citep{Andrews2010}} and $\alpha=10^{-3}$. 
Our fiducial simulation (T1) considers  $q_{\text{I}}=1$ ($M_1=M_2=0.5M_\odot$), $a_{\text{I}}=0.5$~au,  $q_{\text{II}}=1$ ($M_3=1M_\odot$), and $a_{\text{II}}=100$~au. We also perform simulations with different $q_{\text{I}}$ (and thus, different $M_1$ and $M_2$), different $q_{\text{II}}$, and different $a_{\text{I}}$ and $a_{\text{II}}$ in order to test the results dependency on these parameters. The total mass of the inner binary system is always set to $M_{\text{I}}=1M_\odot$, so the disk dynamical time is the same for all models.

\begin{table*}
\caption{Stellar parameters considered for our general simulations of a fixed disk of $M_{\text{d}}=0.05M_\odot$, $R_{\text{c}}=39$~au, $\gamma=1$ and $\alpha=10^{-3}$, and the  resulting disk properties, such as the dissipation timescale $\tau$, the cavity size $R_{\text{c}}$, and the 
truncation radius $R_{\text{t}}$ for the case that considers  viscous accretion by tidal streams with an efficiency of 30\%.
}
\begin{center}
\begin{tabular}{|c|c|c|c|c|c|c|c|c|c|c|c|}
\cline{1-12}
&  &  \multicolumn{7}{ c| }{Stellar Parameters} & \multicolumn{3}{ c| }{Simulation results ($\epsilon=30\%$)}\\ \cline{3-12}
 & Simulation & $M_1$ [$M_\odot$] & $M_2$ [$M_\odot$] & $q_{\text{I}}$ & $a_{\text{I}}$ [au] & $M_3$ [$M_\odot$] & $q_{\text{II}}$ & $a_{\text{II}}$ [au] & $\tau$ [Myr] & $R_{\text{c}}$ [au] & $R_{\text{t}}$ [au]\\
\cline{1-12}
Binary Test Case & B1 & 0.5 & 0.5 & 1 & 0.5 & --- & --- & --- & 4.47 & 1.65 & --- \\
\cline{1-12}
Fiducial Case & T1 & 0.5 & 0.5 & 1 & 0.5 & 1 & 1 & 100 & 3.58 & 1.65 & 34.50  \\
\cline{1-12}
\multicolumn{1}{ |c|  }{\multirow{2}{*}{$q_{\text{I}}$ dependency} } &
\multicolumn{1}{ c| }{T2} & 0.67 & 0.33 & 0.5 & 0.5 & 1 & 1 & 100  & 3.38 & 1.32 & 34.50   \\ \cline{2-12}
\multicolumn{1}{ |c|  }{}                        &  
\multicolumn{1}{ c| }{T3} & 0.91 & 0.09 & 0.1 & 0.5 & 1 & 1 & 100 & 2.80 & 0.85 & 34.50   \\ 
\cline{1-12}
\multicolumn{1}{ |c|  }{\multirow{2}{*}{$q_{\text{II}}$ dependency} } &
\multicolumn{1}{ c| }{T4} & 0.5 & 0.5 & 1 & 0.5 & 0.5 & 0.5 & 100 & 3.80 & 1.65 & 42.30 \\ \cline{2-12}
\multicolumn{1}{ |c|  }{}                        &
\multicolumn{1}{ c| }{T5} & 0.5 & 0.5 & 1 & 0.5 & 0.1 & 0.1 & 100 & 4.17 & 1.65 & 63.5 \\ \cline{1-12}
\multicolumn{1}{ |c|  }{\multirow{2}{*}{$a_{\text{I}}$ dependency} } &
\multicolumn{1}{ c| }{T6} & 0.5 & 0.5 & 1 & 1.5 & 1 & 1 & 100 & 3.57 & 4.82 & 34.50 \\ \cline{2-12}
\multicolumn{1}{ |c|  }{}                        &
\multicolumn{1}{ c| }{T7} & 0.5 & 0.5 & 1 & 2.5 & 1 & 1 & 100 & 3.38 & 7.85 & 34.50 \\ \cline{1-12}
\multicolumn{1}{ |c|  }{\multirow{2}{*}{$a_{\text{II}}$ dependency} } &
\multicolumn{1}{ c| }{T8} & 0.5 & 0.5 & 1 & 0.5 & 1 & 1 & 500 & 4.35 & 1.65 & 168 \\ \cline{2-12}
\multicolumn{1}{ |c|  }{}                        &
\multicolumn{1}{ c| }{T9} & 0.5 & 0.5 & 1 & 0.5 & 1 & 1 & 1000 & 4.36 & 1.65 & 300\\ \cline{1-12}
\end{tabular}
\end{center}
  \label{tab:CI-Generales}
\end{table*}

\section{General Results}
\label{sec:general_results}
In this section we first present the main differences between the disk evolution of our fiducial case in a triple-star system with the same disk in a binary star system, not affected by an external companion. We then describe the results and general characteristics of the simulations of Table \ref{tab:CI-Generales}, where we consider different stellar parameters for triple star systems.

\subsection{Disk evolution: triple-star vs close binary-star systems}
\label{sec:T1-vs-B1}
It has been previously shown that circumbinary disks 
dissipate at a lower rate than circumstellar disks \citep{Alexander2012} 
However, what happens to disks in triple star systems? 
To address this question we performed two simulations: our fiducial case T1 (see Table \ref{tab:CI-Generales}), which represents a circumbinary disk in a triple hierarchical system, and B1, which is exactly the same but ignoring the external star, thus, representing an isolated circumbinary disk. 
We recall that, in terms of our model, the latter is achieved by setting $\Lambda_{\text{II}}=0$.

\subsubsection{Without accretion by tidal streams}
\label{sec:without-VABTS}

In order to focus on how the disk density profile changes with time,
we first do not allow any gas accretion via streams to occur in these simulations. Recall from the end of sec.~\ref{subsec:radial_structure} that, in 1D models, the form of the torque given by eq.~\ref{eq:TorqueI} prevents any mass loss by viscous accretion unless accretion by tidal streams is allowed. Thus, the disk only losses mass by photoevaporation. Since the mass loss rate due to the X-ray photoevaporative winds only depends on the X-ray luminosity of the central object (see eq. \ref{eq:Mdot_hole}), and since T1 and B1 systems, have the same disk mass and the same inner stellar pair, the dissipation timescale for both disks is exactly the same. However, the shapes of the density profiles are quite different. In terms of our model, a gas disk has completely dissipated when its mass in our simulations is lower than $1\times 10^{-6}~\text{M}_{\odot}$. If, as in this case, the disk mass only evolves due to X-ray photoevaporation, the dissipation timescale can be directly estimated without the need of performing simulations, by computing $L_{\text{X}}$ in eq. \ref{eq:XrayL} and then $\dot{M}_{\text{w}}$ in eq. \ref{eq:Mdot_hole}. For both  B1 and T1, $q_{\text{I}}=1$ and $M_1=M_2=0.5M_\odot$, thus we obtain $\dot{M}_{\text{w}}\approx8.9\times10^{-9}~M_\odot$~yr$^{-1}$ and both disks dissipate in $\tau \approx 5.6$~Myr. \\

\begin{figure*}
  \centering
    \includegraphics[angle= 270, width= 0.8\textwidth]{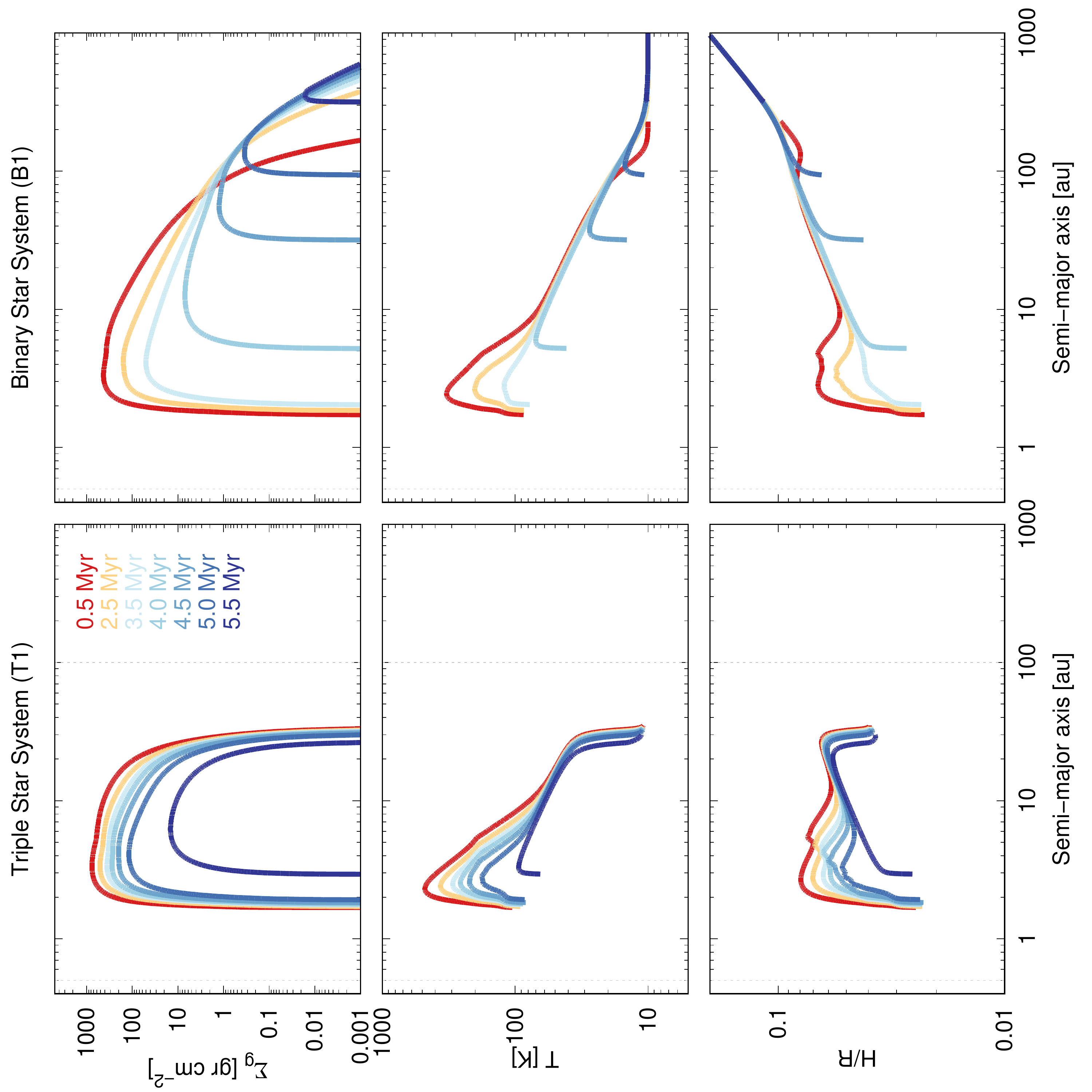}
    \caption{Time evolution for the gas surface density (top), the temperature profiles (middle) and the aspect ratio (bottom) of our fiducial disk (T1) in the hierarchical system (left) and for the same disk in the binary star system (B1), not affected by an external companion. Different color profiles represent the gas disk at different ages. The inner grey vertical dashed lines represent the separation between both stars of the inner binary system, and the outer one (only for T1) the separation between the inner binary and the external star.}
  \label{fig:Comparacion-Triple-Binaria}
\end{figure*}

Figure \ref{fig:Comparacion-Triple-Binaria} shows the time evolution of the gas surface density (top), temperature (middle) and aspect ratio (bottom) profiles for the disk of our fiducial case (T1) in a hierarchical triple star system (left column) compared to the same disk in a binary system (B1) (right column), and, for both cases, without any mass loss by viscous accretion. 

The differences between disk profiles in the triple and binary systems are remarkable. At early times (0.5~Myr of evolution, red profiles), and due to the injection of angular momentum by the inner binary, the inner radius of the gas disk expands, creating an inner cavity of up to $\sim1.65$~au \citep[$\sim 3.30~a_{\text{I}}$ as it is expected from previous studies, e.g.][]{ArtymowiczLubow1994} for both cases, while the outer radius presents different values. In T1, the effect of the outer stellar companion located at 100~au, prevents the disk expansion beyond
$\sim$35~au, which represents $\sim0.35~a_{\text{II}}$ and which is in agreement with previous studies \citep{PapaloizouPringle1977, Pichardo2005}. This situation is different for the circumbinary case, B1, where the disk can freely expand up beyond $\sim$100~au. As a natural consequence of this expansion, the maximum value reached by the gas surface density profile is lower than for the one in the triple star system for about a factor of $\sim$ 0.5 
and this difference increases with time, although the mass for both cases evolve at the same rate (i.e. each comparable, same color, profile between T1 and B1 presents the same amount of mass).

\begin{figure}
    \centering
    \includegraphics[angle= 270, width= 0.99\columnwidth]{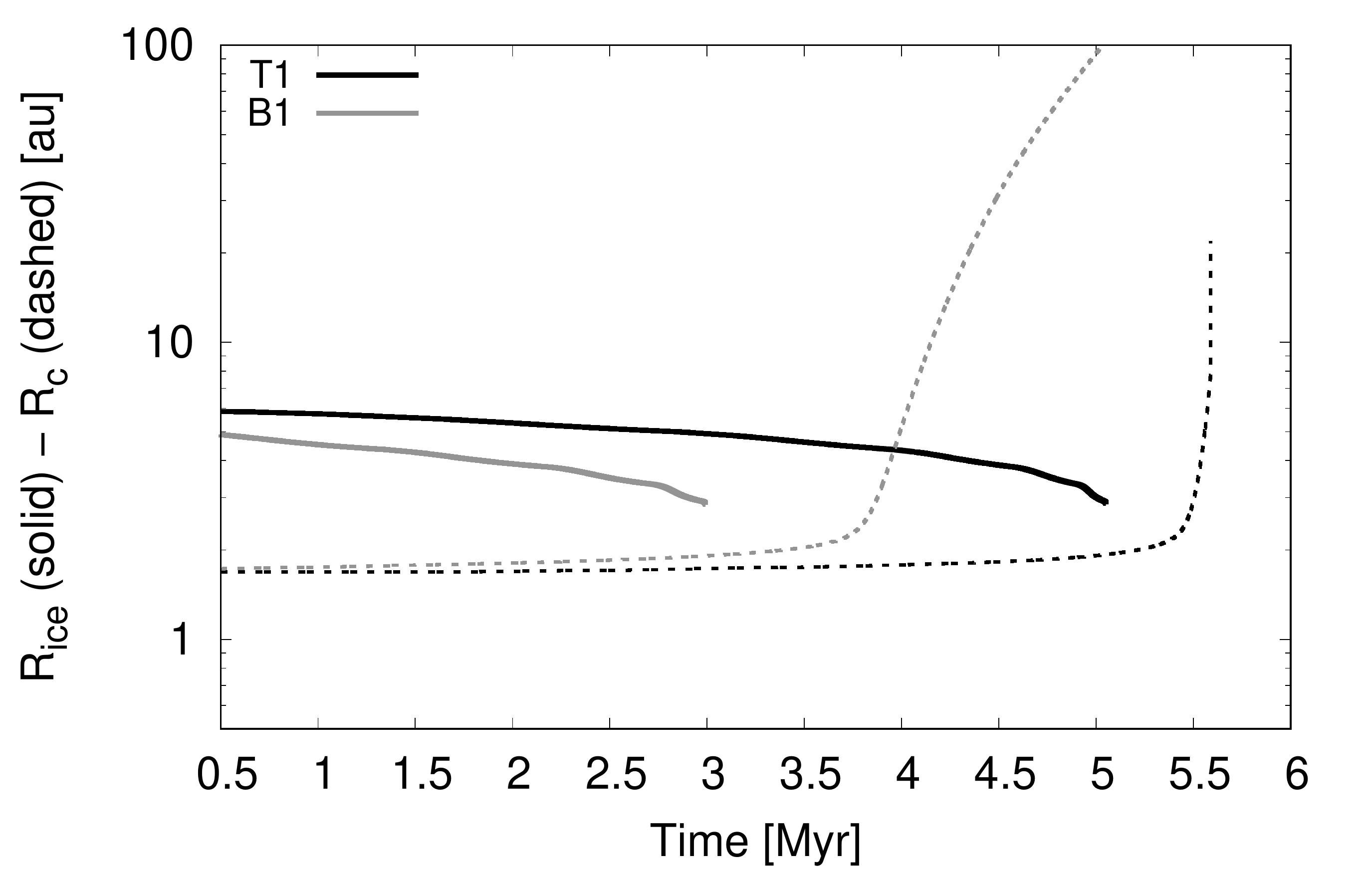}
    \caption{Time evolution of the location of the ice line (solid lines) and the inner radius or cavity size (dashed lines) for T1 and B1.}
  \label{fig:Rhielo-Rhole-T1-B1}
\end{figure}

While the gas density profiles in T1 evolve slowly in a confined region, without significantly changing their shape, in B1 the gas disk is pushed further out generating a growing cavity as a simultaneous consequence of photoevaporation and the disk expansion due to the angular momentum conservation (see Fig. \ref{fig:Rhielo-Rhole-T1-B1} for a comparison between the evolution of the inner cavity sizes). In addition, the gas surface density profiles in B1 decay faster than in T1, thus decreasing the local available mass of gas inside $\sim$35~au. As an example, at 4~Myr the gas surface density at 10~au is $\sim$ 10 gr cm$^{-2}$ for B1 while it is $\sim$ 200 gr cm$^{-2}$ for T1.
This situation could favor gas giant planet formation in triple star systems, in where a protoplanet can spend more time embedded in the gas disk, compared to a circumbinary disks. However, it is important to note that higher gas surface densities can also enhance inward migration of low and intermediate mass planets. 

An interesting point here is that, as a consequence of the higher gas surface density profiles for T1, the mid-plane temperature and aspect ratio profiles of the same disk are higher than the ones of B1. These quantities could play an important role in the solid component evolution of these protoplanetary disks, and thus, in the process of planet formation since higher gas densities also implies higher dust densities.
As an example, the location of the water ice line ($\sim$170~K), beyond which an increment in the solid surface density is expected, is always further out in T1 compared to B1, and it also lasts for longer, as it can be seen in Figure \ref{fig:Rhielo-Rhole-T1-B1}. This situation may lead to the formation of ice cores in T1 for longer timescales and at larger distances than in B1. 
In addition, higher aspect ratios reduce the migration rate of low-mass and intermediate-mass planets \citep[e.g.][]{Paardekooper2011, JimenezMasset2017} and increase the pebble isolation mass \citep[e.g.][]{Lambrechts+2014}. 
These two effects could help planet formation in hierarchical triple star systems more than in binaries. 

On the other hand, we note that while the temperature radial profiles are similar in both cases, i.e. they decrease as we move  away from the central binary, the aspect ratio profiles present more differences. This is particularly important for self-shadowing, although we reiterate here that this effect is not included in the disk thermodynamics of our model.
In any case, self-shadowing would be stronger in the case of the triple than in the circumbinary star system. In the latter case, the decrease in the aspect ratio could generate an effect of self-shadowing until $\sim 10$~au, leading to a cooler disk between $\sim 2$~au and $\sim 10$~au than the previously computed. This effect would be negligible after 2.5~Myr. For the disk in the triple star system, the self-shadowing would affect until $\sim$ 20 au at early times, and persist until $\sim$ 4.5~Myr of the disk evolution. 

\subsubsection{With accretion by tidal streams}
\label{sec:with-VABTS}
In this section, and in order to understand how the differences between the T1 and B1 profiles affect the disk dissipation timescales, we now allow some mass loss due to viscous accretion via streams to occur. For this comparison we adopt an efficiency of 30$\%$, which is an intermediate value between those found by \citet{MacFadyen2008} and \citet{DOrazio2013}.
The general shape and size of the gas, temperature and aspect ratio profiles are very similar to the ones with no mass loss by viscous accretion presented in Figure \ref{fig:Comparacion-Triple-Binaria}. However, they no longer dissipate on the same timescale -- 
the disk in B1 dissipates in 4.47 Myr while the one in T1 dissipates in 3.58 Myr, an $\sim$80\% of the circumbinary disk lifetime. The reason is quite simple and is directly related to the behaviour of the gas disk profiles. As described at the end of section \ref{subsec:radial_structure}, the mass loss rate due to the accretion by tidal streams is proportional to $\Sigma_{\text{g}}$ inside $2R_{\text{c}}$, which is always greater in T1. Thus, the mass removal is higher for the disk in the triple star system though still slow enough to possibly allow planet formation.

\begin{figure}
  \centering
    \includegraphics[angle= 270, width= 0.99\columnwidth]{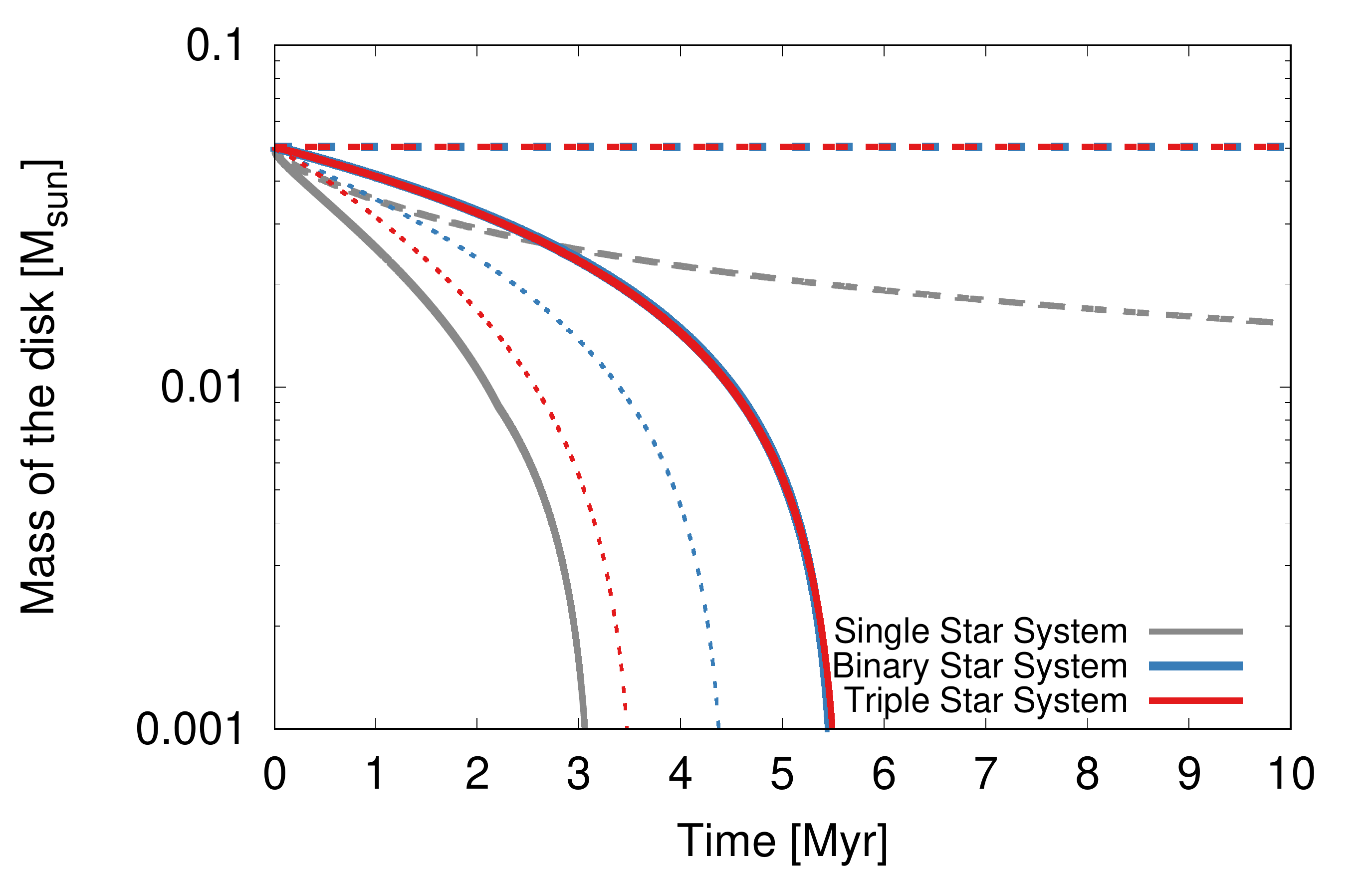}
    \caption{Time evolution of the mass of the disk of our fiducial case T1 in a hierarchical triple star system (red curves) and in the circumbinary case B1 (blue curves). The mass evolution for the corresponding circumstellar disk case is represented in grey. The initial mass of this disk in this case is $M_{\text{d}}=0.05M_\odot$. The solid lines denote the cases in where the photoevaporation due to the inner binary is included, while the long dashed lines represent the cases in where is not. The short dashed lines in red and blue represent the mass of the disks in the triple and binary systems, respectively, with photoevaporation and with an accretion efficiency by streams of 30\%.}
  \label{fig:masas-comparacion}
\end{figure}

Figure \ref{fig:masas-comparacion} shows the time evolution of the mass of the disk in the triple star system T1 in red, the disk in the binary system B1 in blue, and, as a reference, the time evolution of the corresponding circumstellar\footnote{The inner radius of the circumstellar disk is also defined at 0.5~au, as for the case of B1 and T1, in order to be consistent with the comparison. Moreover, to have exactly the same mass loss rate due to the X-ray photoevaporation, the central star is assumed to have the same X-ray luminosity as the inner binary system of B1 and T1.} disk in grey, which for the purposes of our model is computed with $\Lambda_{\text{I}}=\Lambda_{\text{II}}=0$. 
Both the solid and long dashed lines represent cases in which accretion via streams was not considered, and the short dashed lines represent cases in which we consider an accretion efficiency of 30\% by streams. Particularly, the long dashed lines represent cases in which photoevaporation is not included.
In this figure it can be clearly seen that, if photoevaporation is not included, the mass loss by viscous accretion is completely suppressed in B1 and T1, and thus, the disks would remain intact forever.
This is, as mentioned before, a consequence of the form of the torques for 1D models. Then, if we now consider photoevaporation, both the disks in B1 and T1 dissipate on the same timescale because they are only losing mass due to this mechanism. The dissipation timescale changes between B1 and T1, as previously described, when we also allow mass loss by viscous accretion by tidal streams with an efficiency of 30\%. The mass disk evolution is, in this case, represented by the short dashed lines. Although for the triple star system the dissipation timescale is shorter than for the binary case, it is still longer than for the corresponding circumstellar disk (solid grey line). 

\subsection{Stellar parameters dependency}
\label{sec:stellar_dependency}

We now describe the global results of the simulations of Table \ref{tab:CI-Generales} for disks that evolve due to photoevaporation and viscous accretion by tidal streams with an efficiency of $30\%$.
\begin{figure*}
    \centering
    \includegraphics[angle= 270, width= 0.99\textwidth]{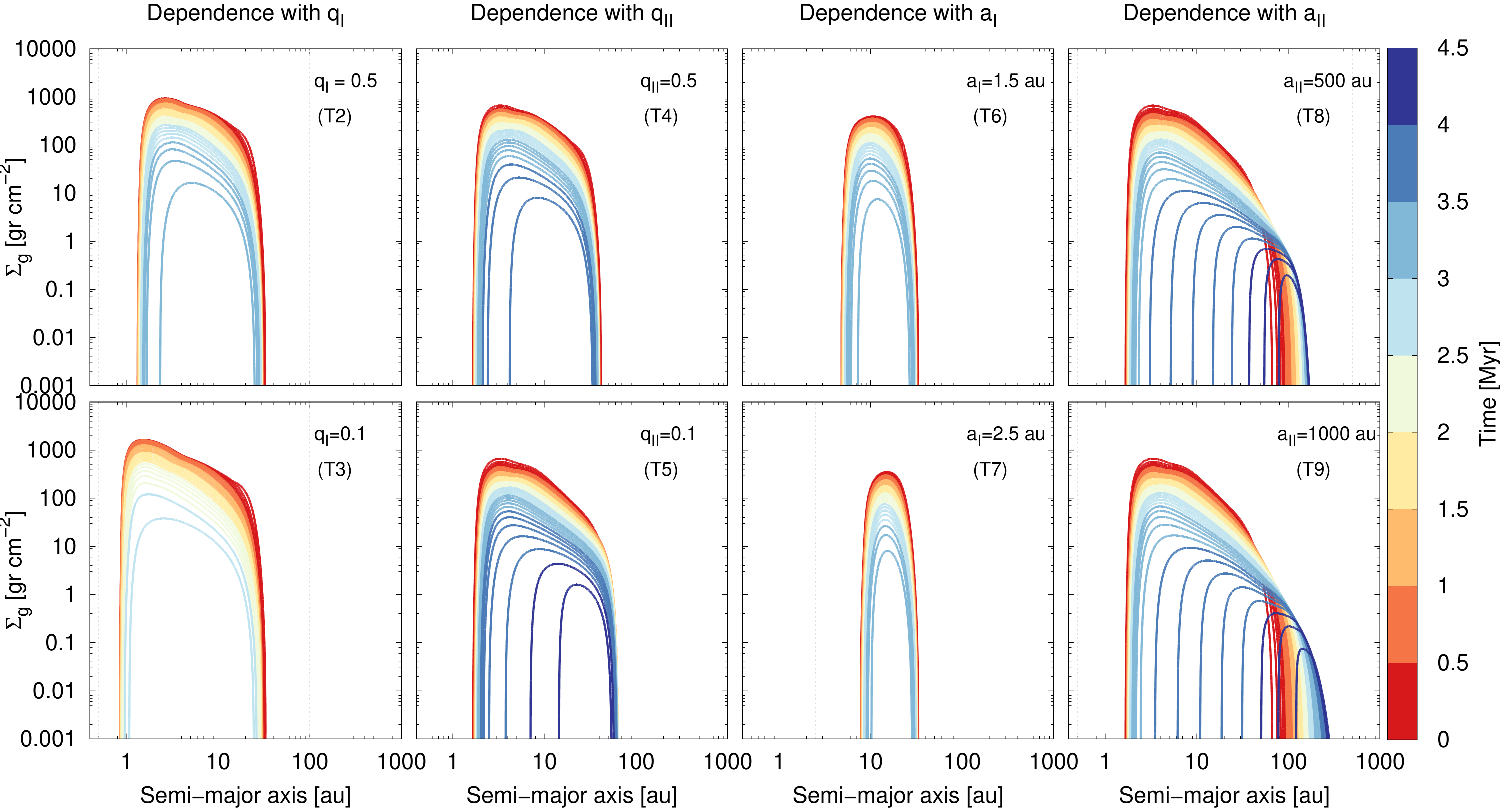}
    \caption{Time evolution, represented by the colorscale, of the gas surface density profiles for simulations T2 to T9 that present different values of $q_{\text{I}}$ (first column), different values of $q_{\text{II}}$ (second column), different values of $a_{\text{I}}$ (third column) and different values of $a_{\text{II}}$ (fourth column). The inner grey vertical dashed lines represent the separation between both stars of the inner binary system, and the outer one the separation between the inner binary and the external star.    } 
  \label{fig:densi-sup-discos-arbitrarios}
\end{figure*}

\subsubsection{Dependence on the stellar mass ratios}
\label{sec:q1q2}
 In this section, we vary the masses of the stars, but always keeping the total mass of the inner stellar binary set to $M_{\text{I}}=1M_\odot$.

We first consider three different values for the mass ratio between components $M_1$ and $M_2$.
In models T1, T2 and T3, we set $q_{\text{I}}=1, 0.5$ and $0.1$, respectively. The mass ratio $q_{\text{I}}$ of the inner binary  plays a key role in the models. On the one hand, $q_{\text{I}}$ enters in eq. \ref{eq:TorqueI}. Thus, the higher the mass ratio, the greater the torque imparted by the binary to the disk, and consequently, the wider the inner cavity and the narrower the disk, since the external star always truncates the disk at $\sim30$~au. The different cavity sizes of the disks can be appreciated in Table \ref{tab:CI-Generales}.
On the other hand, different mass ratios imply different individual masses for the inner system, and, given the non-linear relation of X-ray luminosity with mass, a different total $L_{\text{X}}$. 
Therefore, as $q_{\text{I}}$ grows, $L_{\text{X}}$ decreases, and consequently, it also does the mass loss rate due to photoevaporation (see eq. \ref{eq:Mdot_hole}). Thus, the disk dissipates faster in systems with lower values of $q_{\text{I}}$.
The differences, however, are not very significant and can be found in Table \ref{tab:CI-Generales}. 
The time evolution of the density profiles of T2 and T3 can be seen in the first column of Figure \ref{fig:densi-sup-discos-arbitrarios}. 

To analyze the dependence of the results on $q_{\text{II}}$, we performed two simulations, T4 and T5 with $q_{\text{II}}=0.5$ and $q_{\text{II}}=0.1$, respectively, in addition to our fiducial case T1 with $q_{\text{II}}=1$. Different values of $q_{\text{II}}$ do not play any role in the photoevaporation process since the masses of the inner binary do not change in these simulations and thus, the size of the inner cavity remains the same as for T1. 
What does change in this case is the mass of the external companion. The main effect of $q_{\text{II}}$ is that, as this parameter decreases, the torques exerted by the outer companion to the disk are weaker (see eq. \ref{eq:TorqueII}) and thus, the gas disk is able to expand more (see Table \ref{tab:CI-Generales}). This expansion consequently causes the surface density profiles to decrease, and therefore, there is less mass loss by viscous accretion by streams. The time evolution of the density profiles for these cases can be seen in the second column of Figure \ref{fig:densi-sup-discos-arbitrarios}.

\subsubsection{Dependence on the semi-major axes}
\label{sec:a1a2}

We now test the results dependency on $a_{\text{I}}$ and $a_{\text{II}}$, keeping all the stellar masses fixed as in the fiducial case T1. 
We performed two simulations, T6 and T7, with $a_{\text{I}}=1.5$~au and $a_{\text{I}}=2.5$~au, in addition to our fiducial case T1, with $a_{\text{I}}=0.5$~au. The effect produced by the increase of the inner binary separation is to create a wider inner cavity due to a higher injection of angular momentum to the disk (see eq. \ref{eq:TorqueI}). Consequently, the disk is confined to evolve in an outer and narrower region since the external stellar companion, located at 100~au, truncates the disk at $\sim$35~au. 
Despite the widths of the three cases being different, the dissipation timescales are practically the same, as it can be appreciated in Table \ref{tab:CI-Generales}. 
This happens because the larger the cavity, the smaller the gas surface density at the location in where $\dot{M}_{\text{B}}$ is computed, but the higher the viscosity as it increases at greater distances from the central star. This increment in $\nu$ is a consequence of its proportionally dependence on $c_{\text{s}}$ and ${\text{H}}_{\text{g}}$.
Therefore $\dot{M}_{\text{B}}\propto \nu\Sigma$ remains roughly constant and the accretion rates are very similar for the three cases, while at the same time the photo-evaporation affects the disk equally in the three cases. 

Finally, we run two simulations with different values for $a_{\text{II}}$. Apart from T1 that has $a_{\text{II}}=100$~au, T8 and T9 have $a_{\text{II}}=500$~au and $a_{\text{II}}=1000$~au, respectively. In this case, the effect of increasing the separation between the inner binary and the outer stellar companion allows the disk to evolve in a more extended region.
This expansion of the disk reduces the gas surface densities compared to T1, producing lower mass loss rates due to viscous accretion by streams, which, as already mentioned, is proportional to $\Sigma_{\text{g}}$. Therefore, the dissipation timescales of these disks are longer than the one of T1 ($\sim3.58$~Myr). However the differences between the dissipation timescales of T8 and T9 are insignificant (see Table \ref{tab:CI-Generales}). 
Recall from sec.~\ref{sec:with-VABTS}, that B1, which represents the circumbinary disk case, dissipates in $\sim4.47$~Myr. Thus, disks in triple star systems with sizes larger than $\sim$500~au tend to evolve and dissipate in the same way and on similar timescales as the corresponding circumbinary case. 

\section{The quadruple hierarchical system HD 98800}
\label{sec:HD98800B}

HD 98800 is a hierarchical quadruple stellar system part of the TW Hya association \citep{Soderblom1996,Torres2008, Ducourant2014}. It is formed by two spectroscopic binaries, Aa-Ab (a single-line spectroscopic binary, or SB1) and Ba-Bb (a double-line spectroscopic binary, or SB2) \citep{Torres1995}, 
and by a narrow circumbinary disk that orbits the well characterized system Ba-Bb. The system Ba-Bb is formed by two stars of $\sim$0.6M$_\odot$ and $\sim$0.7M$_\odot$, with a semi-major axis of about 0.5~au, with a high eccentricity estimated to be $\sim$ 0.78 \citep{Boden2005}. The orbit of the system Aa-Ab is instead more uncertain, although it could have a total mass similar 
to that of Ba-Bb, of $\sim$ 1.3M$_\odot$ \citep{Tokovinin1999}, but probably with a lower mass ratio being an SB1 \citep{Torres1995}. 
The estimated separation between both binary systems is $\sim$ 45 au and their mutual orbit presents an eccentricity of about 0.4 \citep{Tokovinin2014}. 
Additionally, the inclination of the outer binary Aa-Ab 
with respect to the circumbinary disk was estimated to be $\approx 88^{\circ}$, and even more, the circumbinary disk is in a polar configuration with respect to Ba-Bb \citep{Kennedy2019}. 

The age of this quadruple system is estimated to be between 7 and 10 Myr \citep{Kastner1997, Prato2001, Barrado2006, Ducourant2014}, which is quite old for a protoplanetary disk. Since moreover there was no detected emission at near-IR wavelengths, the circumbinary disk in HD 98800~B was thought to be a debris disk for many years \citep{Olofsson2012}. Recently, \citet{Ribas2018} were able to resolve the disk with VLA 8.8~mm observations and came to the conclusion that due to its small size, large fractional luminosity, and spectral index consistent with black body emission, the disk should still retain significant amounts of gas, estimated by the authors to be around 5~M$_{\text{J}}$, where M$_{\text{J}}$ is the mass of Jupiter \citep[see Sec.~2 of][for further details]{Ribas2018}.
If this is the case, the question that naturally arises is: How is it possible that the disk still exists in such a complex environment? \citet{Ribas2018} suggested that the disk is long lived because it is only evolving due to photoevaporation, with its  viscous evolution stopped or significantly slowed down by the torques of both binaries. In this section, we apply our model to this particular case to quantitatively test their hypothesis.

\subsection{Modeling the circumbinary disk in HD 98800~B}
\label{sec:HD98800B-Modeling}

As mentioned above, HD 98800 is a quadruple star system. However, in order to model the circumbinary disk around Ba-Bb using {\scriptsize PLANETALP-B}, we consider the outer binary Ab-Aa as a single body.  This is a good approximation for its gravitational effect given the large separation between the binaries, and also the orbit of this component has not been resolved yet.
Moreover, due to the axial symmetry of our model, we need to adopt an important simplification: circular and coplanar orbits for both star systems and the disk. We discuss these simplifications in section \ref{sec:discussion}.  

We model the time evolution of the disk around HD 98800~B by considering stars of masses 0.7M$_\odot$ and 0.6M$_\odot$ for the system Ba-Bb, so their mass ratio is $q_{\text{Ba-Bb}} = 0.857$, and a total mass of 1.3M$_\odot$ for the system Aa-Ab. We consider two different setups for the orbital parameters. 
Setup 1 adopts the observed semi-major axes of the system, namely $a_{\text{Ba-Bb}} = 0.5$~au and $a_{\text{A-B}}= 45$~au, but in a circular and coplanar configuration due to the geometrical limitations of our model. 
For Setup 2 we adopt the extreme values $a_{\text{Ba-Bb}} = 0.9$~au and $a_{\text{A-B}}= 27$~au, that correspond to the maximum and minimum expansion of the Ba-Bb and B-A orbits (apocenter and pericenter), considering their  observed eccentricities of 0.78 and 0.4, respectively, but again, in a circular and coplanar configuration. This range of parameters should capture some of the effects we miss by not having eccentric orbits in the models, allowing us to get realistic estimates on the minimum dissipation timescale of the system. 

Unlike in sec.~\ref{sec:general_results}, we now fix the stellar and orbital parameters of the system and consider different disk parameters such as different values for $M_{\text{d}}$, $\alpha$, and different accretion efficiency fractions $\epsilon$, shown in table \ref{tab:CI-HD98800}. As in our previous simulations of sec.~\ref{sec:general_results}, we keep a fixed exponent for the initial density profile of $\gamma=1$ and a fixed characteristic radius of $R_0=39$~au.

\citet{Kastner2004} was able to measure the X-ray luminosity of HD 98800~B, which is $L^{\text{Ba-Bb}}_{\text{X}} \approx 1.4\times10^{29}$ erg s$^{-1}$. This  value is an order of magnitude lower than the one we would get with our approach $L^{\text{Ba-Bb}}_{\text{X}} = L^{\text{Ba}}_{\text{X}} + L^{\text{Bb}}_{\text{X}} \approx 2.53\times10^{30}$ erg s$^{-1}$ following eqs. \ref{eq:XrayL} and \ref{eq:Lx}, which has important consequences for the disk evolution. 
As explained in Sec.~\ref{sec:T1-vs-B1}, if only X-ray photoevaporation is considered, and no mass loss due to accretion by tidal streams is allowed, the dissipation timescale can be directly estimated without the need of performing simulations. 
Using $L^{\text{Ba-Bb}}_{\text{X}}$ as measured by  \citep{Kastner2004} we obtain disk lifetimes much greater than 10 Myr
for disk masses of $M_{\text{d}}=0.1, 0.05, \text{and } 0.01\,M_\odot$, while using $L_{\text{X}}$ as computed with eqs. \ref{eq:XrayL} and \ref{eq:Lx}, we obtain disk lifetimes of 7.5 Myr, 3.7 Myr, and 0.7 Myr, for the same disk masses. The latter values can be considered as lower limits for the HD 98800~B disk lifetime.

Taking advantage of the fact that this system has a measured X-ray luminosity, we then run the simulations only considering $L^{\text{Ba-Bb}}_{\text{X}}$ as estimated by \citet{Kastner2004}. Due to the uncertainties in the mass loss rates due to stream accretion, we also consider different values for the accretion efficiency $\epsilon$, as shown in Table \ref{tab:CI-HD98800}. In appendix B we analyze the case in where the accretion efficiency fraction is not set as a constant value but as a function of the aspect ratio of the disk, as proposed by \citet{Ragusa2016}. 

\begin{figure*}
  \centering
    \includegraphics[angle= 270, width= 0.98\textwidth]{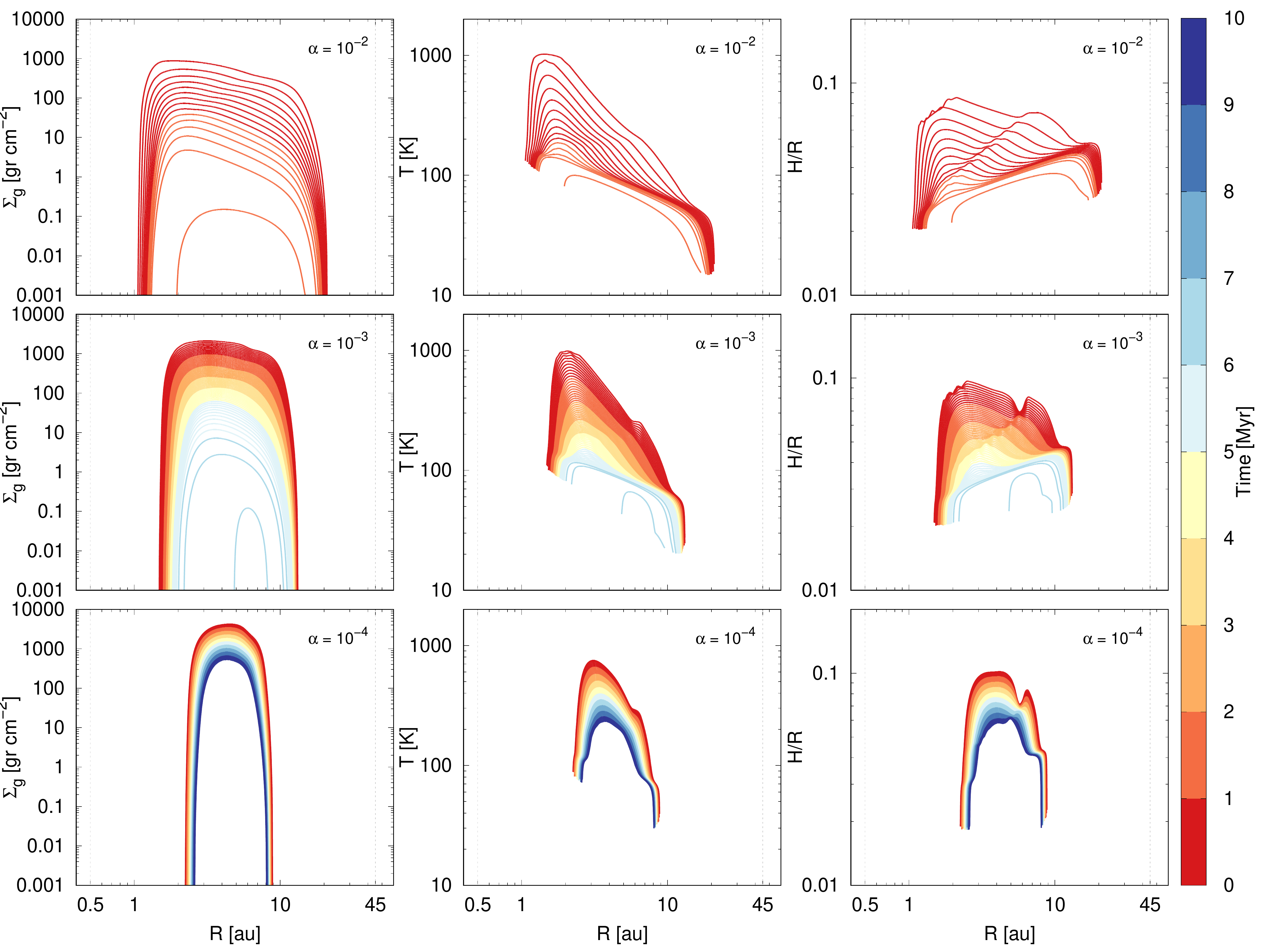}
    \caption{Time evolution of the gas surface density (left column), temperature (middle column) and aspect ratio (right column) for the Sim4 (top row), Sim5 (middle row) and Sim 6(bottom row), considering accretion via streams with an efficiency of 30\% ($\epsilon=0.3$). }
  \label{fig:Sims-4-5-6-HD98800-Setup-1}
\end{figure*}

Figure \ref{fig:Sims-4-5-6-HD98800-Setup-1} shows the time evolution of the gas surface density profiles (left column), temperature profiles (middle column) and aspect ratio profiles (right column) every 0.1 Myr represented by the colorscale, for Sim~4 (top), Sim~5 (middle) and Sim~6 (bottom) of the Setup 1 (see Table \ref{tab:CI-HD98800}) with an accretion efficiency of 30$\%$ ($\epsilon=0.3$). These are the results for a disk of $M_{\text{d}}=0.05M_\odot$ with different $\alpha$ values in the rows. 
As mentioned before, our simulations end when the disk has completely dissipated, defined as a gas disk mass lower than $10^{-6}M_\odot$, or when it reaches 10 Myr.

The viscosity parameter $\alpha$, which is the only difference between the simulations of Fig. \ref{fig:Sims-4-5-6-HD98800-Setup-1}, plays an important role in two different aspects. First, it is responsible for shaping the gas surface density, temperature and disk aspect ratio  profiles, and consequently determining the size and width of the disk, and second, it directly affects the dissipation timescales when some accretion is allowed via streams. 

Regarding sizes, the higher the value of $\alpha$, the wider the disk, which expands in both directions, towards the central binary and towards the outer star. The boundaries for each case, $R_{\text{c}}$ and $R_{\text{t}}$, can be appreciated in Table \ref{tab:CI-HD98800}. These values correspond to the gas disk profile at $\sim$ 0.05 Myr of evolution and measured at $\Sigma_{\text{g}}=10^{-3}$~gr cm$^{-2}$,
and were computed for the case in where no viscous accretion via streams is allowed. However they do not change significantly for the cases in where it is allowed, and thus, for simplicity, we only show in Table \ref{tab:CI-HD98800} the sizes of the disks for the $\epsilon=0$ efficiency cases, both for Setup 1 and 2.  

The time evolution of the gas surface density profiles for disks with $M_{\text{d}}=0.01M_\odot$ and $M_{\text{d}}=0.1M_\odot$ behave in a similar way to the ones described before for $M_{\text{d}}=0.05M_\odot$ but with different dissipation timescales.

\begin{figure}
  \centering
    \includegraphics[angle= 270, width= 0.99\columnwidth]{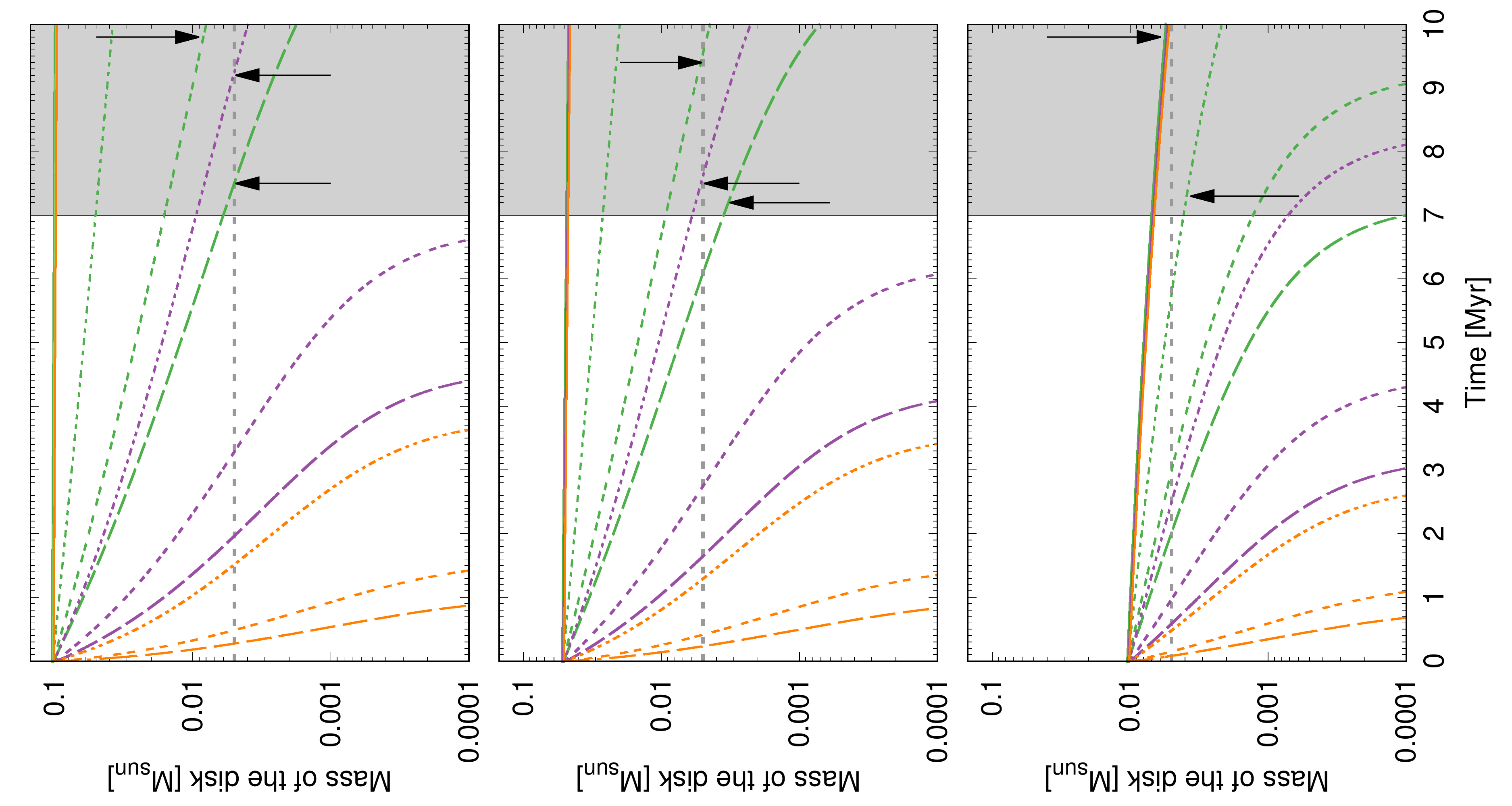}
    \caption{Time evolution of the disk mass in Sim1, Sim2 and Sim3 (bottom panel), Sim4, Sim5 and Sim6 (middle panel), and Sim7, Sim8 and Sim9 (top panel), for $\alpha=0.0001$ (green), $\alpha=0.001$ (purple) and $\alpha=0.01$ (orange), and for accretion efficiencies of 0$\%$ (solid lines), 10$\%$ (short dashed lines), 30$\%$ (middle dashed lines) and 50$\%$ (long dashed lines). The black arrows are highlighting those simulations that fit the estimated mass (horizontal grey dashed line) and age (grey shaded area) of the circumbinary disk HD 98800~B. }
  \label{fig:Evol-Masas}
\end{figure}

Table \ref{tab:CI-HD98800} also shows the dissipation timescales for each of the performed simulations, both for Setup 1 and Setup 2. Those simulations in which the disk dissipates only after the minimum estimated age of HD 98800, 7 Myr, are highlighted in bold both, for the Setup 1 and Setup 2. 
However, only those additionally marked with an $*$ are the ones that also fit the estimated mass $\sim$5M$_{\text{J}}$ of the disk around HD 98800~B  \citep{Ribas2018}.
The latter can be also seen in Fig.~\ref{fig:Evol-Masas}, which shows the time evolution of the mass of the disk for all Setup 1 simulations. The bottom panel shows the results for Sim~1, Sim~2 and Sim~3 with a disk of $M_{\text{d}}=0.01M_\odot$, the middle panel for Sim~4, Sim~5 and Sim~6 with a disk of $M_{\text{d}}=0.05M_\odot$ and the top panel for Sim~7, Sim~8 and Sim~9 with a disk of $M_{\text{d}}=0.1M_\odot$. Green curves represent simulations with $\alpha=0.0001$, purple with $\alpha=0.001$ and orange with $\alpha=0.01$. Solid lines represent simulations with accretion efficiencies of 0$\%$, short dashed lines with 10$\%$, middle dashed lines with 30$\%$  and long dashed lines with 50$\%$. The horizontal grey dashed line is marking the current estimated mass of HD 98800~B \citep{Ribas2018} and the shaded area its probable age. The curves marked with black arrows are the ones highlighted with an $*$ in Table \ref{tab:CI-HD98800}.

From Table \ref{tab:CI-HD98800} and Figure \ref{fig:Evol-Masas} it can be deduced that, despite the complex configuration of HD 98800, and assuming our simplified model for HD 98800, we found several cases that would explain the existence of the massive disk after $\sim$ 7 Myr, even with high mass accretion efficiencies via streams, and with relatively low initial masses. 
However, moderate to low viscosities are needed for the low initial mass cases. It is interesting to remark though, that low viscosity values are the ones that better reproduced some of the ring structures in the DSHARP disks \citep{Dullemond2018}.

As shown by the results in Table \ref{tab:CI-HD98800}, both for Setup 1 and 2, highly viscous disks that are allowed to lose mass  via tidal streams with $\ge 10\%$ efficiency, dissipate in less than 3.81 Myrs even for the most conservative simulations (the most massive disk in Setup 1).

\begin{table*}
\caption{Disk parameters considered for the particular case of HD 98800~B and the corresponding results for Setup 1 and Setup 2. Both setups were computed using the estimated X-ray luminosity of the system by \citet{Kastner2004}. For Setup 1 we adopt $a_{\text{Ba-Bb}} = 0.5$~au and $a_{\text{A-B}}= 45$~au, and for Setup 2, $a_{\text{Ba-Bb}} = 0.9$~au and $a_{\text{A-B}}= 27$~au. 
$\tau$ represents the dissipation timescale of the disk in Myr, $R_{\text{c}}$ is the radius of the inner cavity and $R_{\text{t}}$ the outer truncation radius, both measured at $\Sigma_{\text{g}}=10^{-3}$~gr cm$^{-2}$ and at 0.05 Myr of evolution, and expressed in $a_{\text{Ba-Bb}}$ and $ a_{\text{B-A}}$ units. The resulting timescales greater than 7 Myr are highlighted in boldface and those simulations with $^*$ are the ones that better fit the current mass and age of HD 98800~B.}
\begin{center}
\begin{tabular}{|c|c|c|c|c|c|c|c|c|c|c|c|c|c|c|}
\cline{1-15}
&  \multicolumn{2}{ c| }{Disk } & \multicolumn{6}{ c| }{Setup 1} &  \multicolumn{6}{ c| }{Setup 2}\\
\cline{4-15}
&  \multicolumn{2}{ c| }{Parameters } & \multicolumn{6}{ c| }{Efficiency} & \multicolumn{6}{ c| }{Efficiency}\\ \cline{4-15}
& \multicolumn{2}{ c| }{} & \multicolumn{3}{ c| }{$0\%$} & 
 $10\%$ & $30\%$ & $50\%$ & \multicolumn{3}{ c| }{$0\%$} & 
 $10\%$ & $30\%$ & $50\%$ \\
 \cline{2-15}
 & $M_{\text{d}}$ & $\alpha$ & $R_{\text{c}}$ & $R_{\text{t}}$ & $\tau$ & $\tau$ & $\tau$ & $\tau$ & $R_{\text{c}}$ & $R_{\text{t}}$ & $\tau$ & $\tau$ & $\tau$ & $\tau$ \\
 & [$M_\odot$] & & [$a_{\text{Ba-Bb}}$] & [$a_{\text{A-B}}$] & [Myr] & [Myr] & [Myr] & [Myr] & [$a_{\text{Ba-Bb}}$] & [$a_{\text{A-B}}$] & [Myr] & [Myr] & [Myr] & [Myr] \\
\cline{1-15}
\multicolumn{1}{ |c| }{Sim~1} & 0.01 & $10^{-2}$ & 2.18 & 0.51 & $>$\textbf{10}$^*$ & 2.78 & 1.24 & 0.82 & 2.25 & 0.49 & $>\textbf{10}$ & 1.75 & 0.77 & 0.51 \\ \cline{1-15}
\multicolumn{1}{ |c| }{Sim~2} & 0.01 & $10^{-3}$ & 3.38 & 0.30 & $>$\textbf{10}$^*$ & \textbf{8.31} & 4.49 & 3.21 & 3.15 & 0.31 & $>\textbf{10}$ & 4.46 & 2.17 & 1.50 \\ \cline{1-15}
\multicolumn{1}{ |c| }{Sim~3} & 0.01 & $10^{-4}$ & 5.12 & 0.18 & $>$\textbf{10}$^*$ & $>$\textbf{10}$^*$ & \textbf{9.26} & \textbf{7.19} & 4.21 & 0.22 & $>\textbf{10}$ & \textbf{8.15} & 4.41 & 3.18 \\ \cline{1-15}
\multicolumn{1}{ |c| }{Sim~4} & 0.05 & $10^{-2}$ & 2.06 & 0.55 & $>$\textbf{10} & 3.59 & 1.51 & 0.97 & 2.01 & 0.56 & $>\textbf{10}$ & 2.15 & 0.90 & 0.59 \\ \cline{1-15}
\multicolumn{1}{ |c| }{Sim~5} & 0.05 & $10^{-3}$ & 2.96 & 0.32 & $>$\textbf{10} & $>$\textbf{10}$^*$ & 6.27 & 4.27 & 2.85 & 0.33 & $>\textbf{10}$ & 6.22 & 2.77 & 1.86 \\ \cline{1-15}
\multicolumn{1}{ |c| }{Sim~6} & 0.05 & $10^{-4}$ & 4.48 & 0.21 & $>$\textbf{10} & $>$\textbf{10} & $>$\textbf{10}$^*$ & $>$\textbf{10}$^*$ & 3.89 & 0.24 & $>\textbf{10}$ & $>\textbf{10}$ & 6.17 & 4.26\\ \cline{1-15}
\multicolumn{1}{ |c| }{Sim~7} & 0.1 & $10^{-2}$ & 2.02 & 0.57 & $>$\textbf{10} & 3.81 & 1.57 & 1.01 & 1.96 & 0.59 & $>\textbf{10}$ & 2.55 & 0.94 & 0.61 \\ \cline{1-15}
\multicolumn{1}{ |c| }{Sim~8} & 0.1 & $10^{-3}$ & 2.84 & 0.33 & $>$\textbf{10} & $>$\textbf{10}$^*$ & 6.81 & 4.58 & 2.72 & 0.35 & $>\textbf{10}$ & 6.78 & 2.96 & 1.98 \\ \cline{1-15}
\multicolumn{1}{ |c| }{Sim~9} & 0.1 & $10^{-4}$ & 4.23 & 0.21 & $>$\textbf{10} & $>$\textbf{10} & $>$\textbf{10}$^*$ & $>$\textbf{10}$^*$ & 3.77 & 0.25 & $>\textbf{10}$ & $>\textbf{10}$ & 6.77 & 4.62 \\ 
\cline{1-15}
\end{tabular}
\end{center}
  \label{tab:CI-HD98800}
\end{table*}

\subsection{Could the potential disk around HD 98800~A have already been photoevaporated?}

Given the similar total stellar masses for systems A and B, the lack of a disk in system A is intriguing. We do know, however, that the X-ray luminosity of HD 98800~A is about four times higher than HD 98800~B  \citep{Kastner2004}, which implies a shorter lifetime for a disk around it. 
Here we will test this idea quantitatively with our model.

\begin{table}
\caption{Dissipation timescales for the simulations of Setup 1 for system A.  
Contrary to Table \ref{tab:CI-HD98800}, the resulting timescales shorter than 7 Myr are highlighted in boldface, and those simulations marked with $^\star$ are the ones that, together with the results of Table \ref{tab:CI-HD98800} (and assuming both disks, A and B, initially have the same mass) better fit the current mass and age of HD 98800~B, and the  current non-existence of a disk around system A.}
\begin{center}
\begin{tabular}{|c|c|c|c|c|}
\cline{1-5}
&  \multicolumn{4}{ c| }{Setup 1} \\
\cline{2-5}
&  \multicolumn{4}{ c| }{Efficiency} \\ \cline{2-5}
&  $0\%$ &  $10\%$ & $30\%$ & $50\%$ \\
 \cline{2-5}
 & $\tau$ & $\tau$ & $\tau$ & $\tau$ \\
 & [Myr] & [Myr] & [Myr] & [Myr] \\
\cline{1-5}
\multicolumn{1}{ |c| }{Sim~1} & \textbf{4.20}$^\star$ & --- & --- & ---\\ \cline{1-5}
\multicolumn{1}{ |c| }{Sim~2} & \textbf{4.20}$^\star$ & \textbf{3.88} & --- & ---\\ \cline{1-5}
\multicolumn{1}{ |c| }{Sim~3} & \textbf{4.20}$^\star$ & \textbf{3.87}$^\star$ & \textbf{3.26} & \textbf{2.85} \\ \cline{1-5}
\multicolumn{1}{ |c| }{Sim~4} & $>10$ & --- & --- & --- \\ \cline{1-5}
\multicolumn{1}{ |c| }{Sim~5} & $>10$ & \textbf{6.99}$^\star$ & --- & ---\\ \cline{1-5}
\multicolumn{1}{ |c| }{Sim~6} & $>10$ & $>10$ & 8.06 & \textbf{6.14}$^\star$ \\ \cline{1-5}
\multicolumn{1}{ |c| }{Sim~7} & $>10$ & --- & --- & --- \\ \cline{1-5}
\multicolumn{1}{ |c| }{Sim~8} & $>10$ & 8.53 & --- & --- \\ \cline{1-5}
\multicolumn{1}{ |c| }{Sim~9} & $>10$ & $>10$ & $>10$ & 7.48 \\ 
\cline{1-5}
\end{tabular}
\end{center}
  \label{tab:CI-HD98800A}
\end{table}

We will assume for simplicity that the binary and initial disk in system A have the same parameters as those in B. 
It is important to remark that, because the system Aa-Ab was classified as an SB1 instead of SB2 \citep{Torres1995},  its actual mass ratio should be lower than that of B. As shown in sec.~\ref{sec:q1q2}, a smaller value of $q_{\text{I}}$ leads to a faster disk dispersal.  Therefore, our choice of $q_{\text{A}}=q_{\text{B}}$, even if not very realistic, is conservative as it makes the disk in the A system last for longer.

We compute the time evolution of disks around binary A for orbital Setup I, with $L^{\text{A}}_{\text{X}}= 4L^{\text{B}}_{\text{X}}$, and using the parameters of Table \ref{tab:CI-HD98800} that resulted in dissipation timescales longer than 7 Myr for system B (highlighted in boldface).
Table \ref{tab:CI-HD98800A} presents the dissipation timescales for these new simulations. Those in boldface represent disks that dissipated in less than 7 Myr, and those additionally marked with a $^\star$ are the ones that, together with the results of Table \ref{tab:CI-HD98800} (and assuming both disks, A and B, have initially the same mass) better fit the current mass and age of HD 98800~B, and the current non-existence of a disk around  system A. For the disk around system A, moderate to low mass disks are needed to allow its dissipation timescale in less than  7 Myr, the minimum age estimated for the quadruple system.

\section{Discussion}
\label{sec:discussion}
Here we discuss some important aspects about our model of evolution of disks in triple hierarchical star systems and the possible influence in planet formation.

\subsection{General limitations of our model}

One of the most important hypotheses in our model, consequence of its one-dimensional nature, is the axisymmetry of the disk. This characteristic does not properly reproduce the details of the disk structure, which is clearly non-axisymmetric due to the non-keplerian potential of the central binary. Even if this approximation might not be the most appropriate to describe the shape of the cavity and the inner regions of circumbinary disks \citep{Dunhill2015,Miranda2017,Thun2017,ThunKley2018, Poblete2019}, heavily affected by the inner binary, it works well in the outer regions where the binary potential can be approximated by the potential caused by a point mass. In addition, in the particular case of HD 98800~B, we remark that \citet{Kennedy2019} showed that the disk is largely axisymmetrical.

The second important assumption in our model is the circular and coplanar nature of the orbits of the inner binary, and of the circumbinary disk and external companion with respect to the inner binary. 
On the one hand, binaries on circular orbits produce the most eccentric cavities \citep{Thun2017}, a feature that we cannot reproduce with our 1D+1D model. Although the development of an eccentric cavity affects gas accretion, which results to be variable and modulated by the inner binary orbit, we still take this accretion on average into account.
For our modelling of the disk around HD 98800~B this is not an issue though, as the observed binary is not actually circular. 

On the other hand, although a full coplanar orbital configuration in a hierarchical triple star system could seem unlikely at first sight, a very recent analysis of the system TWA3 that hosts a circumbinary disk, shows that this could be the case \citep{Czekala2021}. Moreover, we do not expect significant changes in our results for the disk around HD 98800~B, which in fact presents a polar configuration with respect to the inner binary system Ba-Bb \citep{Kennedy2019}, since \citet{Franchini2019} showed that for a polar configurations, the main effect on the disk is a reduced inner cavity due to weaker inner torques. 
It is possible though, that the irradiation and photoevaporation mechanisms affecting the circumbinary disk are different due to its polar configuration, since the solid angle intercepted by the disk is different. However, we still do not count with models capable of predicting the effects of irradiation and  photoevaporation on disks from binary star systems, even less if they have polar orbits with respect to their disks. Thus, we could not predict if this configuration would contribute or not to a faster disk dispersal.

\subsection{Effects that could accelerate disk dispersal}

Photoevaporation from the central star is perhaps one of the most important mechanisms affecting mass removal in protoplanetary disks and as shown previously is clearly important for circumbinary disks. Disks in hierarchical triple star systems could have, in addition to the irradiation from the inner binary system, an extra source of X-ray radiation: the external companion. However, as described by \citet{RosottiClarke2018}, the irradiation from the external star could be somehow important only if their separation is lower than the radius at which X-ray heating is effective \citep[which is about 100~au][]{Owen2012}. 
For the particular case of HD 98800~B then, the irradiation from the external binary system Aa-Ab could affect somehow its evolution. However, 3D radiation-hydrodynamics simulations not yet developed would be necessary to evaluate this effect. 

Another mechanism that could also affects disks in general, contributing to a faster disk dispersal, is external photoevaporation from close stellar encounters \citep{Winter2018} or due to OB massive stars nearby \citep{Clarke2007, Anderson2013, winter2020}. \citet{2Shadmehri2018} considered this latter effect in the evolution of circumbinary disks and found that it can significantly affect the disk masses and lifetimes depending on the intensity of the external UV radiation flux.
We plan to include this effect in future works. However it is important to remark that, at least for the Hya association, external photoevaporation does not seem to be
relevant, since even with an estimated age of $\sim$10 Myr still presents disks with detected gas, such as HD 98800~B \citep{Ribas2018} and TWA 3, where $^{12}$CO and $^{13}$CO were detected for the first time \citep{Czekala2021}.

\subsection{Effects on Planet formation}

One of the most important results of this work is that
disks in hierarchical triple star systems can survive for several Myr, even when allowing gas accretion by tidal streams. This situation, together with the fact that surface density values in triples are, in general, higher than the ones of their circumbinary counterparts (see Fig. \ref{fig:Comparacion-Triple-Binaria}), could favor planet formation even more than in circumbinary disks. In addition, the location of the iceline in triple star systems is always beyond the corresponding one in the circumbinary disk case as we shown in sec.~\ref{sec:without-VABTS}. Thus, for equal initial mass disks and dust-to-gas ratio in circumbinary and triple disks, higher surface densities and larger initial locations of the icelines could favor the formation of ice cores at larger distances.  
This comparison between disks in binaries and in triples is analogue to the one between circumstellar and circumbinary disks, where the latter have higher gas surface densities. 
The fact that planet formation in triples, as in binaries, could be favored at larger distances is in line with the idea that planets in circumbinary disks form far away and then migrate inward parking near the stability limit \citep{PierensNelson2007,ThunKley2018}, as it may have happened for the discovered Kepler circumbinary planets. 

A possible problem that circumbinary disks in triple star systems could have to achieve planet formation, is their truncation by the external companion.
Even in the case of having a disk in a binary and in a triple with the same initial mass of gas and solids, the truncation of the disk in the triple avoids the continuous supply of dust and pebbles from the outer region of the disk that could be quickly lost due to radial drift, unlike what happens in circumbinary disks not affected by an external companion. This situation was recently studied by \citet{Zagaria2021}, who found that the dust in circumstellar disks around one of the stars in a binary star system drifts faster than in disks not affected by an outer stellar companion.

Nevertheless, an interesting situation could occur for disks with large $a_{\text{I}}$ (see panels T6 and T7 of Fig.~\ref{fig:densi-sup-discos-arbitrarios}) and  
low viscosities (see bottom row of Fig.~\ref{fig:Sims-4-5-6-HD98800-Setup-1}). In these cases, the gas surface density shows quasi-gaussian radial profiles with a well defined maximum. As it was recently shown by \citet{Guilera2020}, a maximum in the gas surface density is related with a pressure maximum and it represents a preferential location for the dust/pebble trap, planetesimal formation by streaming instability and planet formation by hybrid accretion of pebbles and planetesimals. Moreover, the well defined pressure maximum can act as a planet migration trap, thus, favoring planet formation.

\section{Conclusions}
\label{sec:conclusions}

We present {\scriptsize PLANETALP-B}, a 1D+1D model able to compute the time evolution of the gaseous component of circumbinary disks in triple hierarchical stellar systems in circular and coplanar orbits. Our model considers viscous accretion, irradiation from the central binary and X-ray photoevaporation. This model can also be applied to the study of the evolution of circumbinary disks around close-in binaries, or the evolution of circumstellar disks around one of the stars in a wide binary star system. The general goal of this work is to compare the evolution of circumbinary disks in binaries and triple hierarchical stellar systems and to characterize their particular features and lifetimes. A particular goal is to try to understand how it is possible that some disks in these complex stellar environments survive for longer than typical protoplanetary disks.   

Our main results are briefly summarized as follows
\begin{itemize}
    \item The gas disk profiles in binaries and triples are quite different and these differences are mainly related to the size of the inner cavities and the outer radius of the disks. While disks in triples evolve almost confined in a radial range, disks in binaries migrate and expand radially. 
    \item Disks in triple star systems dissipate faster than their circumbinary counterparts when accretion by tidal streams is considered, as a natural consequence of their higher surface density profiles. However, the difference in their dissipation timescales is only about 20$\%$. 
    \item Disks in triple star systems last for longer timescales than their circumstellar counterparts, even when considering an efficiency for the accretion by tidal streams of 30\%.
    \item Disks in triple star systems dissipate faster for lower values of $q_{\text{I}}$, though the differences are not significant. The opposite happens for lower values of $q_{\text{II}}$.
    \item Higher values of $a_{\text{I}}$ provoke larger inner cavities. However, the mass loss rate due to viscous accretion by tidal streams remains similar as a consequence of a balance produced between the gas surface density and the viscosity. 
    This effect compensates the dissipation timescales which are almost the same for the considered cases.
    \item The main effect of higher values of $a_{\text{II}}$ is that the dissipation timescales are longer.  
    Disks with $a_{\text{II}} > 500$~au tend to dissipate in the same timescales than their corresponding circumbinary disks. 
    \item For the particular analysis of the disk around HD 98800~B, and under certain disk parameters (intermediate to high mass disks and moderate to low viscosities), we are able to explain its estimated age and mass. Simultaneously, we are also able to explain the absence of a disk around the binary system A. 
\end{itemize}
    Thus, our study helps characterizing the main features of disks in hierarchical triple star systems and sheds light on the process of planet formation in these complex environments. 


\acknowledgments

We thank the anonymous referee for his/her enthusiasm about our work and also the editor, Gregory Herczeg, for his useful comments. MPR acknowledges financial support provided by FONDECYT grant 3190336. MPR thanks Richard Alexander for useful discussions on some aspects of the model. MPR, OMG, JC and AB acknowledge support by ANID, -- Millennium Science Initiative Program -- NCN19\_171. OMG is partially support by the PICT 2018-0934 from ANPCyT, Argentina. OMG and M3B are partially supported by the PICT 2016-0053 from ANPCyT, Argentina. This project has received funding from the European Union's Horizon 2020 research and innovation programme under the Marie Sk\l{}odowska-Curie grant agreement No 210021. AB acknowledges support from FONDECYT Regular 1190748.

\bibliography{HD98800}{}

\begin{thebibliography}{}
\expandafter\ifx\csname natexlab\endcsname\relax\def\natexlab#1{#1}\fi
\providecommand{\url}[1]{\href{#1}{#1}}
\providecommand{\dodoi}[1]{doi:~\href{http://doi.org/#1}{\nolinkurl{#1}}}
\providecommand{\doeprint}[1]{\href{http://ascl.net/#1}{\nolinkurl{http://ascl.net/#1}}}
\providecommand{\doarXiv}[1]{\href{https://arxiv.org/abs/#1}{\nolinkurl{https://arxiv.org/abs/#1}}}

\bibitem[{{Akeson} {et~al.}(2019){Akeson}, {Jensen}, {Carpenter}, {Ricci},
  {Laos}, {Nogueira}, \& {Suen-Lewis}}]{Akeson2019}
{Akeson}, R.~L., {Jensen}, E. L.~N., {Carpenter}, J., {et~al.} 2019, \apj, 872,
  158, \dodoi{10.3847/1538-4357/aaff6a}

\bibitem[{{Alexander}(2012)}]{Alexander2012}
{Alexander}, R. 2012, \apjl, 757, L29, \dodoi{10.1088/2041-8205/757/2/L29}

\bibitem[{{Alexander} {et~al.}(2006){Alexander}, {Clarke}, \&
  {Pringle}}]{Alexander.et.al.2006}
{Alexander}, R.~D., {Clarke}, C.~J., \& {Pringle}, J.~E. 2006, \mnras, 369,
  229, \dodoi{10.1111/j.1365-2966.2006.10294.x}

\bibitem[{{Alibert} {et~al.}(2005){Alibert}, {Mordasini}, {Benz}, \&
  {Winisdoerffer}}]{Alibert2005}
{Alibert}, Y., {Mordasini}, C., {Benz}, W., \& {Winisdoerffer}, C. 2005, \aap,
  434, 343, \dodoi{10.1051/0004-6361:20042032}

\bibitem[{{ALMA Partnership} {et~al.}(2015){ALMA Partnership}, {Brogan},
  {P{\'e}rez}, {Hunter}, {Dent}, {Hales}, {Hills}, {Corder}, {Fomalont},
  {Vlahakis}, {Asaki}, {Barkats}, {Hirota}, {Hodge}, {Impellizzeri}, {Kneissl},
  {Liuzzo}, {Lucas}, {Marcelino}, {Matsushita}, {Nakanishi}, {Phillips},
  {Richards}, {Toledo}, {Aladro}, {Broguiere}, {Cortes}, {Cortes}, {Espada},
  {Galarza}, {Garcia-Appadoo}, {Guzman-Ramirez}, {Humphreys}, {Jung}, {Kameno},
  {Laing}, {Leon}, {Marconi}, {Mignano}, {Nikolic}, {Nyman}, {Radiszcz},
  {Remijan}, {Rod{\'o}n}, {Sawada}, {Takahashi}, {Tilanus}, {Vila Vilaro},
  {Watson}, {Wiklind}, {Akiyama}, {Chapillon}, {de Gregorio-Monsalvo}, {Di
  Francesco}, {Gueth}, {Kawamura}, {Lee}, {Nguyen Luong}, {Mangum}, {Pietu},
  {Sanhueza}, {Saigo}, {Takakuwa}, {Ubach}, {van Kempen}, {Wootten},
  {Castro-Carrizo}, {Francke}, {Gallardo}, {Garcia}, {Gonzalez}, {Hill},
  {Kaminski}, {Kurono}, {Liu}, {Lopez}, {Morales}, {Plarre}, {Schieven},
  {Testi}, {Videla}, {Villard}, {Andreani}, {Hibbard}, \&
  {Tatematsu}}]{ALMA2015}
{ALMA Partnership}, {Brogan}, C.~L., {P{\'e}rez}, L.~M., {et~al.} 2015, \apjl,
  808, L3, \dodoi{10.1088/2041-8205/808/1/L3}

\bibitem[{{Alves} {et~al.}(2019){Alves}, {Caselli}, {Girart}, {Segura-Cox},
  {Franco}, {Schmiedeke}, \& {Zhao}}]{Alves2019}
{Alves}, F.~O., {Caselli}, P., {Girart}, J.~M., {et~al.} 2019, Science, 366,
  90, \dodoi{10.1126/science.aaw3491}

\bibitem[{{Anderson} {et~al.}(2013){Anderson}, {Adams}, \&
  {Calvet}}]{Anderson2013}
{Anderson}, K.~R., {Adams}, F.~C., \& {Calvet}, N. 2013, \apj, 774, 9,
  \dodoi{10.1088/0004-637X/774/1/9}

\bibitem[{{Andrews} {et~al.}(2010){Andrews}, {Wilner}, {Hughes}, {Qi}, \&
  {Dullemond}}]{Andrews2010}
{Andrews}, S.~M., {Wilner}, D.~J., {Hughes}, A.~M., {Qi}, C., \& {Dullemond},
  C.~P. 2010, \apj, 723, 1241, \dodoi{10.1088/0004-637X/723/2/1241}

\bibitem[{{Armitage} \& {Natarajan}(2002)}]{ArmitageNatarajan2002}
{Armitage}, P.~J., \& {Natarajan}, P. 2002, \apjl, 567, L9,
  \dodoi{10.1086/339770}

\bibitem[{{Artymowicz} \& {Lubow}(1994)}]{ArtymowiczLubow1994}
{Artymowicz}, P., \& {Lubow}, S.~H. 1994, \apj, 421, 651,
  \dodoi{10.1086/173679}

\bibitem[{{Artymowicz} \& {Lubow}(1996)}]{ArtymowiczLubow1996}
---. 1996, \apjl, 467, L77, \dodoi{10.1086/310200}

\bibitem[{{Baraffe} {et~al.}(2015){Baraffe}, {Homeier}, {Allard}, \&
  {Chabrier}}]{Baraffe2015}
{Baraffe}, I., {Homeier}, D., {Allard}, F., \& {Chabrier}, G. 2015, \aap, 577,
  A42, \dodoi{10.1051/0004-6361/201425481}

\bibitem[{{Barrado Y Navascu{\'e}s}(2006)}]{Barrado2006}
{Barrado Y Navascu{\'e}s}, D. 2006, \aap, 459, 511,
  \dodoi{10.1051/0004-6361:20065717}

\bibitem[{{Bate}(2018)}]{Bate2018}
{Bate}, M.~R. 2018, \mnras, 475, 5618, \dodoi{10.1093/mnras/sty169}

\bibitem[{{Boden} {et~al.}(2005){Boden}, {Sargent}, {Akeson}, {Carpenter},
  {Torres}, {Latham}, {Soderblom}, {Nelan}, {Franz}, \&
  {Wasserman}}]{Boden2005}
{Boden}, A.~F., {Sargent}, A.~I., {Akeson}, R.~L., {et~al.} 2005, \apj, 635,
  442, \dodoi{10.1086/497328}

\bibitem[{{Bonavita} \& {Desidera}(2020)}]{BonavitaDesidera2020}
{Bonavita}, M., \& {Desidera}, S. 2020, Galaxies, 8, 16,
  \dodoi{10.3390/galaxies8010016}

\bibitem[{{Cieza} {et~al.}(2009){Cieza}, {Padgett}, {Allen}, {McCabe},
  {Brooke}, {Carey}, {Chapman}, {Fukagawa}, {Huard}, {Noriga-Crespo},
  {Peterson}, \& {Rebull}}]{Cieza2009}
{Cieza}, L.~A., {Padgett}, D.~L., {Allen}, L.~E., {et~al.} 2009, \apjl, 696,
  L84, \dodoi{10.1088/0004-637X/696/1/L84}

\bibitem[{{Cieza} {et~al.}(2019){Cieza}, {Ru{\'\i}z-Rodr{\'\i}guez}, {Hales},
  {Casassus}, {P{\'e}rez}, {Gonzalez-Ruilova}, {C{\'a}novas}, {Williams},
  {Zurlo}, {Ansdell}, {Avenhaus}, {Bayo}, {Bertrang}, {Christiaens}, {Dent},
  {Ferrero}, {Gamen}, {Olofsson}, {Orcajo}, {Pe{\~n}a Ram{\'\i}rez},
  {Principe}, {Schreiber}, \& {van der Plas}}]{Cieza2019}
{Cieza}, L.~A., {Ru{\'\i}z-Rodr{\'\i}guez}, D., {Hales}, A., {et~al.} 2019,
  \mnras, 482, 698, \dodoi{10.1093/mnras/sty2653}

\bibitem[{{Clarke}(2007)}]{Clarke2007}
{Clarke}, C.~J. 2007, \mnras, 376, 1350,
  \dodoi{10.1111/j.1365-2966.2007.11547.x}

\bibitem[{{Correia} {et~al.}(2006){Correia}, {Zinnecker}, {Ratzka}, \&
  {Sterzik}}]{Correia2006}
{Correia}, S., {Zinnecker}, H., {Ratzka}, T., \& {Sterzik}, M.~F. 2006, \aap,
  459, 909, \dodoi{10.1051/0004-6361:20065545}

\bibitem[{{Cox} {et~al.}(2017){Cox}, {Harris}, {Looney}, {Chiang}, {Chandler},
  {Kratter}, {Li}, {Perez}, \& {Tobin}}]{Cox2017}
{Cox}, E.~G., {Harris}, R.~J., {Looney}, L.~W., {et~al.} 2017, \apj, 851, 83,
  \dodoi{10.3847/1538-4357/aa97e2}

\bibitem[{{Cuadra} {et~al.}(2009){Cuadra}, {Armitage}, {Alexander}, \&
  {Begelman}}]{Cuadra2009}
{Cuadra}, J., {Armitage}, P.~J., {Alexander}, R.~D., \& {Begelman}, M.~C. 2009,
  \mnras, 393, 1423, \dodoi{10.1111/j.1365-2966.2008.14147.x}

\bibitem[{{Czekala} {et~al.}(2021){Czekala}, {Ribas}, {Cuello}, {Chiang},
  {Mac{\'\i}as}, {Duch{\^e}ne}, {Andrews}, \& {Espaillat}}]{Czekala2021}
{Czekala}, I., {Ribas}, {\'A}., {Cuello}, N., {et~al.} 2021, arXiv e-prints,
  arXiv:2102.11875.
\newblock \doarXiv{2102.11875}

\bibitem[{{del Valle} \& {Escala}(2012)}]{DelValle2012}
{del Valle}, L., \& {Escala}, A. 2012, \apj, 761, 31,
  \dodoi{10.1088/0004-637X/761/1/31}

\bibitem[{{D'Orazio} {et~al.}(2013){D'Orazio}, {Haiman}, \&
  {MacFadyen}}]{DOrazio2013}
{D'Orazio}, D.~J., {Haiman}, Z., \& {MacFadyen}, A. 2013, \mnras, 436, 2997,
  \dodoi{10.1093/mnras/stt1787}

\bibitem[{{Duch{\^e}ne} \& {Kraus}(2013)}]{DucheneKrauss2013}
{Duch{\^e}ne}, G., \& {Kraus}, A. 2013, \araa, 51, 269,
  \dodoi{10.1146/annurev-astro-081710-102602}

\bibitem[{{Ducourant} {et~al.}(2014){Ducourant}, {Teixeira}, {Galli}, {Le
  Campion}, {Krone-Martins}, {Zuckerman}, {Chauvin}, \& {Song}}]{Ducourant2014}
{Ducourant}, C., {Teixeira}, R., {Galli}, P.~A.~B., {et~al.} 2014, \aap, 563,
  A121, \dodoi{10.1051/0004-6361/201322075}

\bibitem[{{Dullemond} {et~al.}(2018){Dullemond}, {Birnstiel}, {Huang},
  {Kurtovic}, {Andrews}, {Guzm{\'a}n}, {P{\'e}rez}, {Isella}, {Zhu}, {Benisty},
  {Wilner}, {Bai}, {Carpenter}, {Zhang}, \& {Ricci}}]{Dullemond2018}
{Dullemond}, C.~P., {Birnstiel}, T., {Huang}, J., {et~al.} 2018, \apjl, 869,
  L46, \dodoi{10.3847/2041-8213/aaf742}

\bibitem[{{Dunhill} {et~al.}(2015){Dunhill}, {Cuadra}, \&
  {Dougados}}]{Dunhill2015}
{Dunhill}, A.~C., {Cuadra}, J., \& {Dougados}, C. 2015, \mnras, 448, 3545,
  \dodoi{10.1093/mnras/stv284}

\bibitem[{{Ercolano} \& {Owen}(2010)}]{ErcolanoOwen2010}
{Ercolano}, B., \& {Owen}, J.~E. 2010, \mnras, 406, 1553,
  \dodoi{10.1111/j.1365-2966.2010.16798.x}

\bibitem[{{Farris} {et~al.}(2014){Farris}, {Duffell}, {MacFadyen}, \&
  {Haiman}}]{Farris2014}
{Farris}, B.~D., {Duffell}, P., {MacFadyen}, A.~I., \& {Haiman}, Z. 2014, \apj,
  783, 134, \dodoi{10.1088/0004-637X/783/2/134}

\bibitem[{{Fedele} {et~al.}(2010){Fedele}, {van den Ancker}, {Henning},
  {Jayawardhana}, \& {Oliveira}}]{Fedele2010}
{Fedele}, D., {van den Ancker}, M.~E., {Henning}, T., {Jayawardhana}, R., \&
  {Oliveira}, J.~M. 2010, \aap, 510, A72, \dodoi{10.1051/0004-6361/200912810}

\bibitem[{{Fontecilla} {et~al.}(2019){Fontecilla}, {Haiman}, \&
  {Cuadra}}]{Fontecilla2019}
{Fontecilla}, C., {Haiman}, Z., \& {Cuadra}, J. 2019, \mnras, 482, 4383,
  \dodoi{10.1093/mnras/sty2972}

\bibitem[{{Franchini} {et~al.}(2019){Franchini}, {Lubow}, \&
  {Martin}}]{Franchini2019}
{Franchini}, A., {Lubow}, S.~H., \& {Martin}, R.~G. 2019, \apjl, 880, L18,
  \dodoi{10.3847/2041-8213/ab2fd8}

\bibitem[{{Gorti} {et~al.}(2009){Gorti}, {Dullemond}, \&
  {Hollenbach}}]{Gorti2009}
{Gorti}, U., {Dullemond}, C.~P., \& {Hollenbach}, D. 2009, \apj, 705, 1237,
  \dodoi{10.1088/0004-637X/705/2/1237}

\bibitem[{{Guilera} {et~al.}(2019){Guilera}, {Cuello}, {Montesinos}, {Miller
  Bertolami}, {Ronco}, {Cuadra}, \& {Masset}}]{Guilera2019}
{Guilera}, O.~M., {Cuello}, N., {Montesinos}, M., {et~al.} 2019, \mnras, 486,
  5690, \dodoi{10.1093/mnras/stz1158}

\bibitem[{{Guilera} {et~al.}(2017){Guilera}, {Miller Bertolami}, \&
  {Ronco}}]{Guilera2017}
{Guilera}, O.~M., {Miller Bertolami}, M.~M., \& {Ronco}, M.~P. 2017, \mnras,
  471, L16, \dodoi{10.1093/mnrasl/slx095}

\bibitem[{{Guilera} {et~al.}(2020){Guilera}, {S{\'a}ndor}, {Ronco},
  {Venturini}, \& {Miller Bertolami}}]{Guilera2020}
{Guilera}, O.~M., {S{\'a}ndor}, Z., {Ronco}, M.~P., {Venturini}, J., \& {Miller
  Bertolami}, M.~M. 2020, arXiv e-prints, arXiv:2005.10868.
\newblock \doarXiv{2005.10868}

\bibitem[{{Hirsh} {et~al.}(2020){Hirsh}, {Price}, {Gonzalez},
  {Ubeira-Gabellini}, \& {Ragusa}}]{Hirsh2020}
{Hirsh}, K., {Price}, D.~J., {Gonzalez}, J.-F., {Ubeira-Gabellini}, M.~G., \&
  {Ragusa}, E. 2020, \mnras, 498, 2936, \dodoi{10.1093/mnras/staa2536}

\bibitem[{{Jim{\'e}nez} \& {Masset}(2017)}]{JimenezMasset2017}
{Jim{\'e}nez}, M.~A., \& {Masset}, F.~S. 2017, \mnras, 471, 4917,
  \dodoi{10.1093/mnras/stx1946}

\bibitem[{{Kastner} {et~al.}(2004){Kastner}, {Huenemoerder}, {Schulz},
  {Canizares}, {Li}, \& {Weintraub}}]{Kastner2004}
{Kastner}, J.~H., {Huenemoerder}, D.~P., {Schulz}, N.~S., {et~al.} 2004, \apjl,
  605, L49, \dodoi{10.1086/420769}

\bibitem[{{Kastner} {et~al.}(1997){Kastner}, {Zuckerman}, {Weintraub}, \&
  {Forveille}}]{Kastner1997}
{Kastner}, J.~H., {Zuckerman}, B., {Weintraub}, D.~A., \& {Forveille}, T. 1997,
  Science, 277, 67, \dodoi{10.1126/science.277.5322.67}

\bibitem[{{Kennedy} {et~al.}(2019){Kennedy}, {Matr{\`a}}, {Facchini}, {Milli},
  {Pani{\'c}}, {Price}, {Wilner}, {Wyatt}, \& {Yelverton}}]{Kennedy2019}
{Kennedy}, G.~M., {Matr{\`a}}, L., {Facchini}, S., {et~al.} 2019, Nature
  Astronomy, 3, 230, \dodoi{10.1038/s41550-018-0667-x}

\bibitem[{{Kraus} {et~al.}(2012){Kraus}, {Ireland}, {Hillenbrand}, \&
  {Martinache}}]{Kraus2012}
{Kraus}, A.~L., {Ireland}, M.~J., {Hillenbrand}, L.~A., \& {Martinache}, F.
  2012, \apj, 745, 19, \dodoi{10.1088/0004-637X/745/1/19}

\bibitem[{{Kraus} {et~al.}(2020){Kraus}, {Kreplin}, {Young}, {Bate}, {Monnier},
  {Harries}, {Avenhaus}, {Kluska}, {Laws}, {Rich}, {Willson}, {Aarnio},
  {Adams}, {Andrews}, {Anugu}, {Bae}, {ten Brummelaar}, {Calvet}, {Cur{\'e}},
  {Davies}, {Ennis}, {Espaillat}, {Gardner}, {Hartmann}, {Hinkley}, {Labdon},
  {Lanthermann}, {LeBouquin}, {Schaefer}, {Setterholm}, {Wilner}, \&
  {Zhu}}]{Kraus2020}
{Kraus}, S., {Kreplin}, A., {Young}, A.~K., {et~al.} 2020, Science, 369, 1233,
  \dodoi{10.1126/science.aba4633}

\bibitem[{{Kunitomo} {et~al.}(2021){Kunitomo}, {Ida}, {Takeuchi}, {Pani{\'c}},
  {Miley}, \& {Suzuki}}]{Kunitomo2021}
{Kunitomo}, M., {Ida}, S., {Takeuchi}, T., {et~al.} 2021, \apj, 909, 109,
  \dodoi{10.3847/1538-4357/abdb2a}

\bibitem[{{Lambrechts} {et~al.}(2014){Lambrechts}, {Johansen}, \&
  {Morbidelli}}]{Lambrechts+2014}
{Lambrechts}, M., {Johansen}, A., \& {Morbidelli}, A. 2014, \aap, 572, A35,
  \dodoi{10.1051/0004-6361/201423814}

\bibitem[{{Lodato} {et~al.}(2009){Lodato}, {Nayakshin}, {King}, \&
  {Pringle}}]{Lodato2009}
{Lodato}, G., {Nayakshin}, S., {King}, A.~R., \& {Pringle}, J.~E. 2009, \mnras,
  398, 1392, \dodoi{10.1111/j.1365-2966.2009.15179.x}

\bibitem[{{MacFadyen} \& {Milosavljevi{\'c}}(2008)}]{MacFadyen2008}
{MacFadyen}, A.~I., \& {Milosavljevi{\'c}}, M. 2008, \apj, 672, 83,
  \dodoi{10.1086/523869}

\bibitem[{{Manara} {et~al.}(2019){Manara}, {Tazzari}, {Long}, {Herczeg},
  {Lodato}, {Rota}, {Cazzoletti}, {van der Plas}, {Pinilla}, {Dipierro},
  {Edwards}, {Harsono}, {Johnstone}, {Liu}, {Menard}, {Nisini}, {Ragusa},
  {Boehler}, \& {Cabrit}}]{Manara2019}
{Manara}, C.~F., {Tazzari}, M., {Long}, F., {et~al.} 2019, \aap, 628, A95,
  \dodoi{10.1051/0004-6361/201935964}

\bibitem[{{Martin}(2018)}]{Martin2018}
{Martin}, D.~V. 2018, {Populations of Planets in Multiple Star Systems}, ed.
  H.~J. {Deeg} \& J.~A. {Belmonte}, 156, \dodoi{10.1007/978-3-319-55333-7_156}

\bibitem[{{Marzari} \& {Thebault}(2019)}]{MarzariThebault2019}
{Marzari}, F., \& {Thebault}, P. 2019, Galaxies, 7, 84,
  \dodoi{10.3390/galaxies7040084}

\bibitem[{{Migaszewski}(2015)}]{Migaszewski2015}
{Migaszewski}, C. 2015, \mnras, 453, 1632, \dodoi{10.1093/mnras/stv1739}

\bibitem[{{Miranda} {et~al.}(2017){Miranda}, {Mu{\~n}oz}, \&
  {Lai}}]{Miranda2017}
{Miranda}, R., {Mu{\~n}oz}, D.~J., \& {Lai}, D. 2017, \mnras, 466, 1170,
  \dodoi{10.1093/mnras/stw3189}

\bibitem[{{Olofsson} {et~al.}(2012){Olofsson}, {Juh{\'a}sz}, {Henning},
  {Mutschke}, {Tamanai}, {Mo{\'o}r}, \& {{\'A}brah{\'a}m}}]{Olofsson2012}
{Olofsson}, J., {Juh{\'a}sz}, A., {Henning}, T., {et~al.} 2012, \aap, 542, A90,
  \dodoi{10.1051/0004-6361/201118735}

\bibitem[{{Owen} {et~al.}(2012){Owen}, {Clarke}, \& {Ercolano}}]{Owen2012}
{Owen}, J.~E., {Clarke}, C.~J., \& {Ercolano}, B. 2012, \mnras, 422, 1880,
  \dodoi{10.1111/j.1365-2966.2011.20337.x}

\bibitem[{{Owen} {et~al.}(2011){Owen}, {Ercolano}, \& {Clarke}}]{Owen2011}
{Owen}, J.~E., {Ercolano}, B., \& {Clarke}, C.~J. 2011, \mnras, 412, 13,
  \dodoi{10.1111/j.1365-2966.2010.17818.x}

\bibitem[{{Paardekooper} {et~al.}(2011){Paardekooper}, {Baruteau}, \&
  {Kley}}]{Paardekooper2011}
{Paardekooper}, S.-J., {Baruteau}, C., \& {Kley}, W. 2011, \mnras, 410, 293,
  \dodoi{10.1111/j.1365-2966.2010.17442.x}

\bibitem[{{Papaloizou} \& {Pringle}(1977)}]{PapaloizouPringle1977}
{Papaloizou}, J., \& {Pringle}, J.~E. 1977, \mnras, 181, 441,
  \dodoi{10.1093/mnras/181.3.441}

\bibitem[{{Petrovich} \& {Rafikov}(2012)}]{PetrovichRafikov2012}
{Petrovich}, C., \& {Rafikov}, R.~R. 2012, \apj, 758, 33,
  \dodoi{10.1088/0004-637X/758/1/33}

\bibitem[{{Pfalzner} {et~al.}(2014){Pfalzner}, {Steinhausen}, \&
  {Menten}}]{Pfalzner2014}
{Pfalzner}, S., {Steinhausen}, M., \& {Menten}, K. 2014, \apjl, 793, L34,
  \dodoi{10.1088/2041-8205/793/2/L34}

\bibitem[{{Pichardo} {et~al.}(2005){Pichardo}, {Sparke}, \&
  {Aguilar}}]{Pichardo2005}
{Pichardo}, B., {Sparke}, L.~S., \& {Aguilar}, L.~A. 2005, \mnras, 359, 521,
  \dodoi{10.1111/j.1365-2966.2005.08905.x}

\bibitem[{{Picogna} {et~al.}(2019){Picogna}, {Ercolano}, {Owen}, \&
  {Weber}}]{Picogna2019}
{Picogna}, G., {Ercolano}, B., {Owen}, J.~E., \& {Weber}, M.~L. 2019, \mnras,
  487, 691, \dodoi{10.1093/mnras/stz1166}

\bibitem[{{Pierens} \& {Nelson}(2007)}]{PierensNelson2007}
{Pierens}, A., \& {Nelson}, R.~P. 2007, \aap, 472, 993,
  \dodoi{10.1051/0004-6361:20077659}

\bibitem[{{Poblete} {et~al.}(2019){Poblete}, {Cuello}, \&
  {Cuadra}}]{Poblete2019}
{Poblete}, P.~P., {Cuello}, N., \& {Cuadra}, J. 2019, \mnras, 489, 2204,
  \dodoi{10.1093/mnras/stz2297}

\bibitem[{{Prato} {et~al.}(2001){Prato}, {Ghez}, {Pi{\~n}a}, {Telesco},
  {Fisher}, {Wizinowich}, {Lai}, {Acton}, \& {Stomski}}]{Prato2001}
{Prato}, L., {Ghez}, A.~M., {Pi{\~n}a}, R.~K., {et~al.} 2001, \apj, 549, 590,
  \dodoi{10.1086/319061}

\bibitem[{{Preibisch} {et~al.}(2005){Preibisch}, {Kim}, {Favata}, {Feigelson},
  {Flaccomio}, {Getman}, {Micela}, {Sciortino}, {Stassun}, {Stelzer}, \&
  {Zinnecker}}]{Preibisch2005}
{Preibisch}, T., {Kim}, Y.-C., {Favata}, F., {et~al.} 2005, \apjs, 160, 401,
  \dodoi{10.1086/432891}

\bibitem[{{Pringle}(1981)}]{Pringle1981}
{Pringle}, J.~E. 1981, \araa, 19, 137,
  \dodoi{10.1146/annurev.aa.19.090181.001033}

\bibitem[{{Rafikov}(2016)}]{Rafikov2016}
{Rafikov}, R.~R. 2016, \apj, 827, 111, \dodoi{10.3847/0004-637X/827/2/111}

\bibitem[{{Rafikov} \& {Petrovich}(2012)}]{RafikovPetrovich2012}
{Rafikov}, R.~R., \& {Petrovich}, C. 2012, \apj, 747, 24,
  \dodoi{10.1088/0004-637X/747/1/24}

\bibitem[{{Ragusa} {et~al.}(2020){Ragusa}, {Alexander}, {Calcino}, {Hirsh}, \&
  {Price}}]{Ragusa2020}
{Ragusa}, E., {Alexander}, R., {Calcino}, J., {Hirsh}, K., \& {Price}, D.~J.
  2020, \mnras, 499, 3362, \dodoi{10.1093/mnras/staa2954}

\bibitem[{{Ragusa} {et~al.}(2016){Ragusa}, {Lodato}, \& {Price}}]{Ragusa2016}
{Ragusa}, E., {Lodato}, G., \& {Price}, D.~J. 2016, \mnras, 460, 1243,
  \dodoi{10.1093/mnras/stw1081}

\bibitem[{{Reipurth} {et~al.}(2014){Reipurth}, {Clarke}, {Boss}, {Goodwin},
  {Rodr{\'\i}guez}, {Stassun}, {Tokovinin}, \& {Zinnecker}}]{Reipurth2014}
{Reipurth}, B., {Clarke}, C.~J., {Boss}, A.~P., {et~al.} 2014, in Protostars
  and Planets VI, ed. H.~{Beuther}, R.~S. {Klessen}, C.~P. {Dullemond}, \&
  T.~{Henning}, 267, \dodoi{10.2458/azu_uapress_9780816531240-ch012}

\bibitem[{{Ribas} {et~al.}(2018){Ribas}, {Mac{\'\i}as}, {Espaillat}, \&
  {Duch{\^e}ne}}]{Ribas2018}
{Ribas}, {\'A}., {Mac{\'\i}as}, E., {Espaillat}, C.~C., \& {Duch{\^e}ne}, G.
  2018, \apj, 865, 77, \dodoi{10.3847/1538-4357/aad81b}

\bibitem[{{Ronco} {et~al.}(2017){Ronco}, {Guilera}, \& {de
  El{\'{\i}}a}}]{Ronco+2017}
{Ronco}, M.~P., {Guilera}, O.~M., \& {de El{\'{\i}}a}, G.~C. 2017, \mnras, 471,
  2753, \dodoi{10.1093/mnras/stx1746}

\bibitem[{{Rosotti} \& {Clarke}(2018)}]{RosottiClarke2018}
{Rosotti}, G.~P., \& {Clarke}, C.~J. 2018, \mnras, 473, 5630,
  \dodoi{10.1093/mnras/stx2769}

\bibitem[{{Schwarz} {et~al.}(2016){Schwarz}, {Funk}, {Zechner}, \&
  {Bazs{\'o}}}]{Schwarz2016}
{Schwarz}, R., {Funk}, B., {Zechner}, R., \& {Bazs{\'o}}, {\'A}. 2016, \mnras,
  460, 3598, \dodoi{10.1093/mnras/stw1218}

\bibitem[{{Shadmehri} {et~al.}(2018){Shadmehri}, {Ghoreyshi}, \&
  {Alipour}}]{2Shadmehri2018}
{Shadmehri}, M., {Ghoreyshi}, S.~M., \& {Alipour}, N. 2018, \apj, 867, 41,
  \dodoi{10.3847/1538-4357/aae2b5}

\bibitem[{{Shakura} \& {Sunyaev}(1973)}]{Shakura1973}
{Shakura}, N.~I., \& {Sunyaev}, R.~A. 1973, \aap, 24, 337

\bibitem[{{Soderblom} {et~al.}(1996){Soderblom}, {Henry}, {Shetrone}, {Jones},
  \& {Saar}}]{Soderblom1996}
{Soderblom}, D.~R., {Henry}, T.~J., {Shetrone}, M.~D., {Jones}, B.~F., \&
  {Saar}, S.~H. 1996, \apj, 460, 984, \dodoi{10.1086/177026}

\bibitem[{{Tang} {et~al.}(2017){Tang}, {MacFadyen}, \& {Haiman}}]{Tang2017}
{Tang}, Y., {MacFadyen}, A., \& {Haiman}, Z. 2017, \mnras, 469, 4258,
  \dodoi{10.1093/mnras/stx1130}

\bibitem[{{Tazzari} \& {Lodato}(2015)}]{TazzariLodato2015}
{Tazzari}, M., \& {Lodato}, G. 2015, \mnras, 449, 1118,
  \dodoi{10.1093/mnras/stv352}

\bibitem[{{Thun} \& {Kley}(2018)}]{ThunKley2018}
{Thun}, D., \& {Kley}, W. 2018, \aap, 616, A47,
  \dodoi{10.1051/0004-6361/201832804}

\bibitem[{{Thun} {et~al.}(2017){Thun}, {Kley}, \& {Picogna}}]{Thun2017}
{Thun}, D., {Kley}, W., \& {Picogna}, G. 2017, \aap, 604, A102,
  \dodoi{10.1051/0004-6361/201730666}

\bibitem[{{Tobin} {et~al.}(2016){Tobin}, {Kratter}, {Persson}, {Looney},
  {Dunham}, {Segura-Cox}, {Li}, {Chandler}, {Sadavoy}, {Harris}, {Melis}, \&
  {P{\'e}rez}}]{Tobin2016}
{Tobin}, J.~J., {Kratter}, K.~M., {Persson}, M.~V., {et~al.} 2016, \nat, 538,
  483, \dodoi{10.1038/nature20094}

\bibitem[{{Tokovinin} {et~al.}(2014){Tokovinin}, {Mason}, \&
  {Hartkopf}}]{Tokovinin2014}
{Tokovinin}, A., {Mason}, B.~D., \& {Hartkopf}, W.~I. 2014, \aj, 147, 123,
  \dodoi{10.1088/0004-6256/147/5/123}

\bibitem[{{Tokovinin}(1999)}]{Tokovinin1999}
{Tokovinin}, A.~A. 1999, Astronomy Letters, 25, 669

\bibitem[{{Torres} {et~al.}(2008){Torres}, {Quast}, {Melo}, \&
  {Sterzik}}]{Torres2008}
{Torres}, C.~A.~O., {Quast}, G.~R., {Melo}, C.~H.~F., \& {Sterzik}, M.~F. 2008,
  {Young Nearby Loose Associations}, ed. B.~{Reipurth}, Vol.~5, 757

\bibitem[{{Torres} {et~al.}(1995){Torres}, {Stefanik}, {Latham}, \&
  {Mazeh}}]{Torres1995}
{Torres}, G., {Stefanik}, R.~P., {Latham}, D.~W., \& {Mazeh}, T. 1995, \apj,
  452, 870, \dodoi{10.1086/176355}

\bibitem[{{Vartanyan} {et~al.}(2016){Vartanyan}, {Garmilla}, \&
  {Rafikov}}]{Vartanyan2016}
{Vartanyan}, D., {Garmilla}, J.~A., \& {Rafikov}, R.~R. 2016, \apj, 816, 94,
  \dodoi{10.3847/0004-637X/816/2/94}

\bibitem[{{Venturini} {et~al.}(2020{\natexlab{a}}){Venturini}, {Guilera},
  {Ronco}, \& {Mordasini}}]{venturini2020SE}
{Venturini}, J., {Guilera}, O.~M., {Ronco}, M.~P., \& {Mordasini}, C.
  2020{\natexlab{a}}, arXiv e-prints, arXiv:2008.05497.
\newblock \doarXiv{2008.05497}

\bibitem[{{Venturini} {et~al.}(2020{\natexlab{b}}){Venturini}, {Ronco}, \&
  {Guilera}}]{Venturini2020}
{Venturini}, J., {Ronco}, M.~P., \& {Guilera}, O.~M. 2020{\natexlab{b}},
  Submmited to Space Science Review

\bibitem[{{Winter} {et~al.}(2018){Winter}, {Clarke}, {Rosotti}, {Ih},
  {Facchini}, \& {Haworth}}]{Winter2018}
{Winter}, A.~J., {Clarke}, C.~J., {Rosotti}, G., {et~al.} 2018, \mnras, 478,
  2700, \dodoi{10.1093/mnras/sty984}

\bibitem[{{Winter} {et~al.}(2020){Winter}, {Kruijssen}, {Longmore}, \&
  {Chevance}}]{winter2020}
{Winter}, A.~J., {Kruijssen}, J.~M.~D., {Longmore}, S.~N., \& {Chevance}, M.
  2020, \nat, 586, 528, \dodoi{10.1038/s41586-020-2800-0}

\bibitem[{{Zagaria} {et~al.}(2021){Zagaria}, {Rosotti}, \&
  {Lodato}}]{Zagaria2021}
{Zagaria}, F., {Rosotti}, G.~P., \& {Lodato}, G. 2021, \mnras, 504, 2235,
  \dodoi{10.1093/mnras/stab985}

\bibitem[{{Zurlo} {et~al.}(2020){Zurlo}, {Cieza}, {P{\'e}rez}, {Christiaens},
  {Williams}, {Guidi}, {C{\'a}novas}, {Casassus}, {Hales}, {Principe},
  {Ru{\'\i}z-Rodr{\'\i}guez}, \& {Fernandez-Figueroa}}]{Zurlo2020}
{Zurlo}, A., {Cieza}, L.~A., {P{\'e}rez}, S., {et~al.} 2020, \mnras, 496, 5089,
  \dodoi{10.1093/mnras/staa1886}

\end{thebibliography}
\bibliographystyle{aasjournal}



\appendix

\section{Validation of our 1D+1D model: vertical structure and gas evolution for circumbinary disks}

In order to validate our 1D+1D model for the vertical structure and the time evolution of the gas in circumbinary disks added in {\scriptsize PLANETALP-B}, we compare our results against those of \citet{Vartanyan2016}, only for their case with $M_{\text{d}}=0.1M_\odot$ (their Fig. 5 and 6).

We first compute the vertical structure of the disk using the same stellar and disk parameters they considered but computing the irradiation of the disk following our Eq. \ref{eq:Irrad1} and \ref{eq:Irrad2}. The disk and stellar parameters are  $a_{\text{I}}=0.2$~au, $\alpha=0.01$, $M_{\text{d}}=0.1M_\odot$, $L_{\text{I}}=2L_\odot$, $q_{\text{I}}=1$ and central binary of total mass $M_{\text{I}}=1M_\odot$.

As it was also noticed by \citet{2Shadmehri2018}, the binary torque equation in \citet{Vartanyan2016} is a bit different from the classical one from \citet{ArmitageNatarajan2002}. In order to properly compare our model with that of \citet{Vartanyan2016}, we use here the same expression they used, with a factor $f=2\times10^{-3}$ (see their Eq. 3). 

We then compute the time evolution of the gaseous component by solving Eq. \ref{eq:evol_gas} without photoevaporation from the central star, this is, $\dot{\Sigma}_{\text{w}}(R)=0$. The initial profile used by \citet{Vartanyan2016} is a narrow initial peak distribution in the shape of a ring given by 
\begin{eqnarray}
  \Sigma_{\text{g}} &=& \frac{M_{\text{d}}}{(2\pi)^{\frac{3}{2}}R_0\sigma_{\text{r}}} {\text{exp}}\left({-\dfrac{(R-R_0)^2}{2\sigma^2_{\text{r}}}}\right), 
  \label{eq1-sec2-0}
\end{eqnarray}
where $R_0=20$~au, is the initial radius of the disk, and $\sigma_{\text{r}}=2$~au, is its width.

\begin{figure}
  \centering
    \includegraphics[angle= 270, width= 0.8\textwidth]{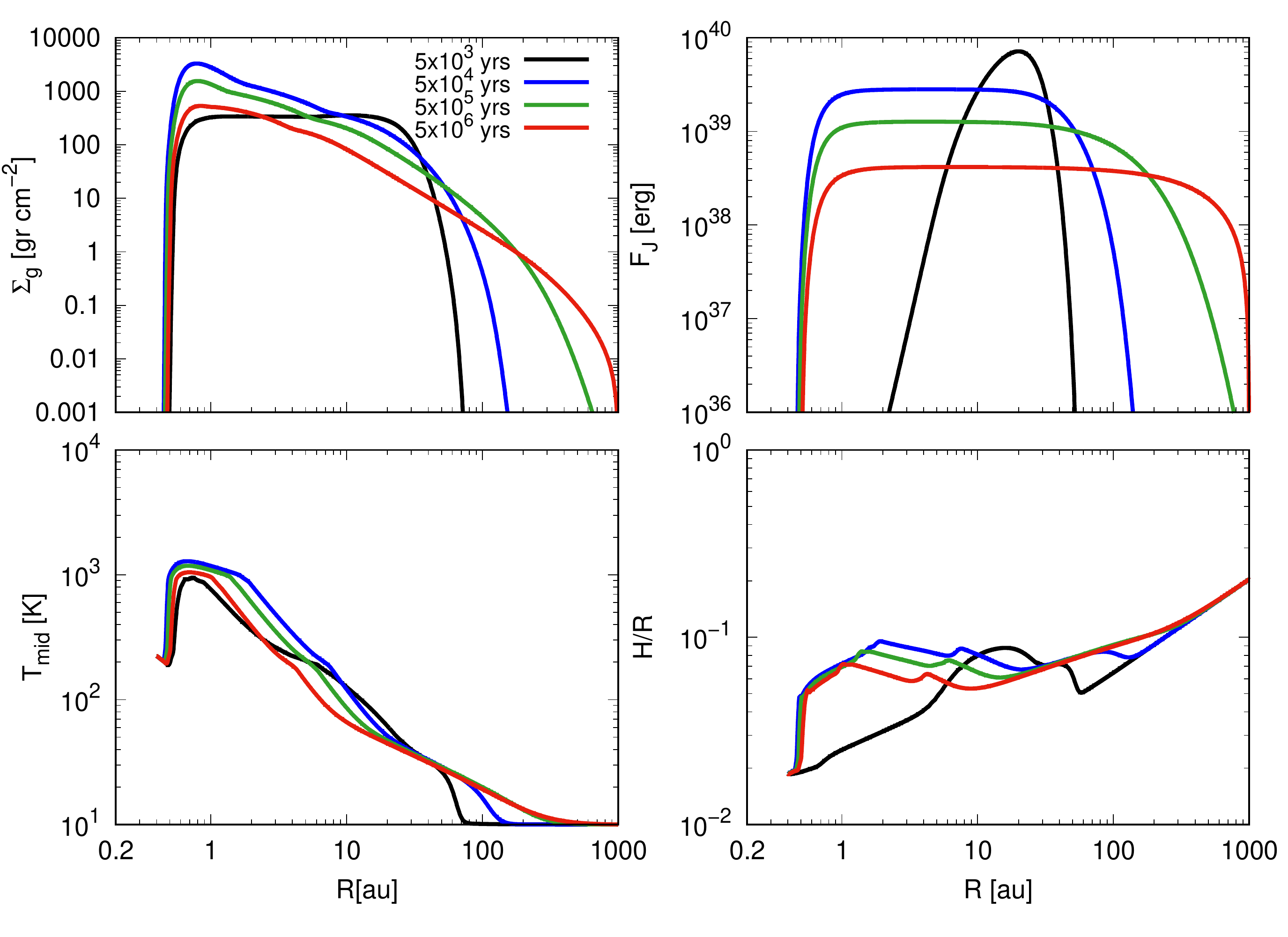}
    \caption{Time evolution of the gas surface density (top left panel), Flux (top right panel), mid plane temperature (bottom left) and aspect ratio (bottom right) for the comparison case with \citet{Vartanyan2016} (see their Figs. 5 and 6): $M_{\text{I}}=1M_\odot$, $q=1$, $a_{\text{I}}=0.2$~au, $r_{\text{in}}=0.2$~au, $\alpha=0.01$, and $M_{\text{d}}=0.1M_\odot$. }
  \label{fig:Comparacion-Vartanyan}
\end{figure}

Figure \ref{fig:Comparacion-Vartanyan} shows the gas surface density (top left panel), viscous angular momentum Flux profiles computed as $F_{\text{J}}=3\pi\nu\Sigma l$, with $l=\Omega R^2$ the specific angular momentum (top right panel), mid plane temperature (bottom left panel) and the aspect ratio (bottom right) for the comparison case and at the same times showed by \citet{Vartanyan2016}. The flux profiles can be compared with their Fig. 3 and the gas and temperature profiles can be compared with Fig. 6 of their work. The aspect ratio in \citet{Vartanyan2016} is only shown for a disk of $M_{\text{d}}=0.05M_\odot$ (see their Fig.11), however it presents very similar characteristics to our case of $M_{\text{d}}=0.1M_\odot$. 

As in \citet{Vartanyan2016}, our disks would also be prone to self-shadowing caused by the tidal and viscous heating enhancement of the mid plane temperature and thus the aspect ratio of the inner regions of the disk. This effect can be clearly appreciated in the bottom left and right panels of Fig. \ref{fig:Comparacion-Vartanyan}, in where beyond $\sim20$~au, H/R presents a flared structure, but between $\sim1$~au and $\sim$20~au H/R increases as R gets smaller. This effect could change the temperature in this region, which could be lower in fact. As pointed out by \citet{Vartanyan2016}, taking into account the self-shadowing in 1D+1D models is extremely hard and is completely out of the scope of this work. 

In general, our results look very similar to those of \citet{Vartanyan2016} despite some minor differences mainly related to the fact that our irradiation model is different. We particularly reproduce quite well the inner cavity of the disk, that expands almost up to $\sim$ 0.5~au, between 2--3 times the binary separation, and the very early plateau distribution in the flux. 

\section{Accretion efficiency as a function of the aspect ratio}

The evolution of circumbinary disks, studied by means of 1D models, predicts that viscous accretion of material is not possible. As mentioned before, this result is due to the particular way of modeling the viscous torques. Instead, 2D and 3D simulations have shown that viscous accretion can happen  via  gas  tidal  streams inside the cavity with accretion rates that could be similar or higher than those of single star systems. However, this scenario can drastically change when the disks become thin. \citet{Ragusa2016} found that when the aspect ratio of the disk ($\text{H}/\text{R}$) is $\gtrsim$ 0.1
the accretion efficiency is not affected by the tidal torques of the inner binary, however, for thinner disks with $\text{H}/\text{R} <$ 0.1, the accretion efficiency linearly decays as
\begin{equation}
  \epsilon(\text{H}/\text{R}) \approx \begin{cases}
      10(\text{H}/\text{R}) & \text{if} ~~\text{H}/\text{R} \leq 0.1 \\
      1 & \text{if} ~~\text{H}/\text{R} > 0.1.
      \label{eq:efficiency_aspect_ratio}
    \end{cases}
\end{equation}
Here we compute again the time evolution of the circumbinary disk of Sim~6 - Setup I (see Table \ref{tab:CI-HD98800}) but instead of considering a fixed value for the accretion efficiency as we did in sec.~\ref{sec:HD98800B-Modeling}, we follow eq. \ref{eq:efficiency_aspect_ratio}.

\begin{figure}
  \centering
    \includegraphics[angle= 270, width= 0.5\textwidth]{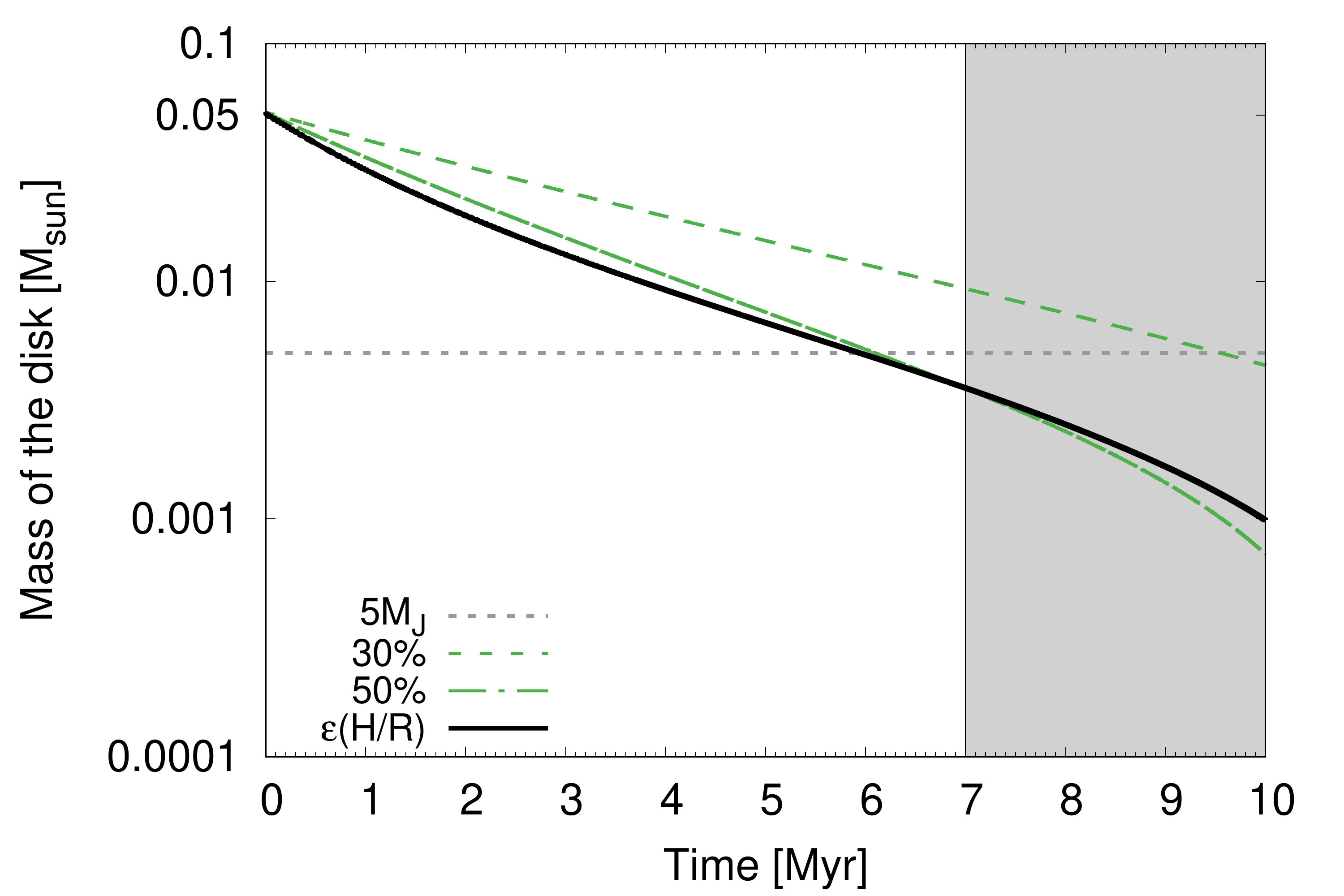}
    \caption{Time evolution of the disk of Sim~6 - Setup I (see Table \ref{tab:CI-HD98800}). The green short and long dashed curves are computed with a fixed value for the accretion efficiency of the 30\% and 50\%, respectively. The solid black curve is computed with an accretion efficiency fraction that depends on the aspect ratio of the disk by following eq. \ref{eq:efficiency_aspect_ratio}. }
  \label{fig:AppB}
\end{figure}

Figure \ref{fig:AppB} shows the time evolution of the mass of the disk of Sim 6 (the solid black line) computed following eq. \ref{eq:efficiency_aspect_ratio} and compared to the cases computed with a fixed value of the accretion efficiency of 30$\%$ ($\epsilon=0.3$) and of 50$\%$ ($\epsilon=0.5$), in short and long dashed lines, respectively. As it can be seen, the more realistic approach derived by \citet{Ragusa2016} is in good agreement with an accretion efficiency of $\sim$50\% regardless of the aspect ratio of the disk.

\end{document}